\documentclass[twocolumn,twocolappendix,tighten]{aastex631}

\usepackage{xspace}
\usepackage{amsmath, amssymb}
\usepackage{ulem}

\hypersetup{linkcolor=red,citecolor=orange,filecolor=cyan,urlcolor=magenta}

\def\lesssim{\mathrel{\hbox{\rlap{\hbox{\lower4pt\hbox{$\sim$}}}\hbox{$<$}}}}
\def\gtrsim{\mathrel{\hbox{\rlap{\hbox{\lower4pt\hbox{$\sim$}}}\hbox{$>$}}}}

\newcommand{\bea}{\begin{eqnarray}}
\newcommand{\eea}{\end{eqnarray}}

\newcommand{\dF}{{^{^*}\!\!F}}

\newcommand{\bF}{{\bf F}}
\newcommand{\bU}{{\bf U}}

\newcommand{\del}{{\partial}}

\newcommand{\prim}{{{\mathbf{P}}}}

\newcommand{\beq}[1]{\begin{equation} #1 \end{equation}}
\newcommand{\beqa}[1]{\begin{eqnarray} #1 \end{eqnarray}}
\newcommand{\deriv}[2]{\frac{ d #1 }{ d #2 }}

\newcommand{\harm}{\xspace{\sc Harm3d}\xspace}

\newcommand{\rpmax}{R_{p}}
\newcommand{\rin}{r_\mathrm{in}}
\newcommand{\rout}{r_\mathrm{out}}
\newcommand{\rdiskin}{R_\mathrm{disk \, in}}
\newcommand{\rdiskout}{R_\mathrm{disk \, out}}

\newcommand{\elump}{e_\mathrm{lump}}
\newcommand{\phase}{\varphi}
\newcommand{\lnorm}[2]{||#2||_{#1}}
\newcommand{\Omegabin}{\Omega_\mathrm{bin}}
\newcommand{\omegalump}{\omega_\mathrm{lump}}
\newcommand{\phaselump}{\phase_\mathrm{lump}}
\newcommand{\phasebin}{\phase_\mathrm{bin}}
\newcommand{\tbin}{t_\mathrm{bin}}
\newcommand{\tend}{t_\mathrm{end}}

\newcommand{\tlump}{t_\mathrm{lump}}
\newcommand{\Tlump}{T_\mathrm{lump}}
\newcommand{\rlump}{r_\mathrm{lump}}
\newcommand{\drlump}{\Delta \rlump}
\newcommand{\dtlump}{\Delta \tlump}
\newcommand{\dtadv}{\Delta t_\mathrm{adv}}
\newcommand{\critlump}{C_\mathrm{lump}}
\newcommand{\tdiss}{t_\mathrm{diss}}

\newcommand{\shellavg}[1]{\langle #1 \rangle}
\newcommand{\shellint}[1]{\left\{ #1 \right\}}
\newcommand{\timesmooth}[2]{\overline{#1}\left(#2\right)}
\newcommand{\massavg}[1]{{\shellavg{#1}}_\rho}
\newcommand{\mdotvsr}[1]{\overline{\dot{M}(t,#1)}}
\newcommand{\mbh}{\mathrm{M}}

\newcommand{\xx    }[1]{x^{\left(#1\right)}}
\newcommand{\QQ    }[1]{Q^{\left(#1\right)}}
\newcommand{\QQa   }[1]{\langle \QQ{#1} \rangle_{\rho}}
\newcommand{\xone  }{\xx{1}}
\newcommand{\xtwo  }{\xx{2}}
\newcommand{\xthree}{\xx{3}}

\newcommand{\plotdir}{./}

\def\lambdabar{%
\relax
\bgroup
\def\@tempa{\hbox{\raise.73\ht0
\hbox to0pt{\kern.25\wd0\vrule width.5\wd0
height.1pt depth.1pt\hss}\box0}}%
\mathchoice{\setbox0\hbox{$\displaystyle\lambda$}\@tempa}%
{\setbox0\hbox{$\textstyle\lambda$}\@tempa}%
{\setbox0\hbox{$\scriptstyle\lambda$}\@tempa}%
{\setbox0\hbox{$\scriptscriptstyle\lambda$}\@tempa}%
\egroup
}

\newcommand{\runse}{\xspace{RunSE}\xspace}
\newcommand{\qone}{\runse}
\newcommand{\qtwo}{\xspace{Run$_{q=1/2}$}\xspace}
\newcommand{\qfive}{\xspace{Run$_{q=1/5}$}\xspace}
\newcommand{\qten}{\xspace{Run$_{q=1/10}$}\xspace}
\newcommand{\origrun}{\runse}
\newcommand{\lrgrun}{\xspace{Run$_{lrg}$}\xspace}
\newcommand{\medrun}{\xspace{Run$_{med}$}\xspace}
\newcommand{\injectrun}{\xspace{Run$_{inj}$}\xspace}
\newcommand{\massratioruns}{\xspace{mass ratio series}\xspace}
\newcommand{\massratiorunscap}{\xspace{Mass Ratio Series}\xspace}
\newcommand{\magfluxruns}{\xspace{magnetic flux series}\xspace}
\newcommand{\magfluxrunscap}{\xspace{Magnetic Flux Series}\xspace}

\newcommand{\massratiolist}{\xspace{\qone, \qtwo, \qfive, \qten}\xspace}
\newcommand{\magfluxlist}{\xspace{\origrun, \medrun, \lrgrun, \injectrun}\xspace}

\submitjournal{ApJ}

\shorttitle{Mass-ratio and Magnetic Flux-Dependence of Circumbinary Disks}
\shortauthors{Noble, et al. (2021)}

\begin{document}

\title{Mass-ratio and Magnetic Flux-Dependence of Modulated Accretion from Circumbinary Disks}

\correspondingauthor{Scott~C.~Noble}
\email{scott.c.noble@nasa.gov}
\author[0000-0003-3547-8306]{Scott~C.~Noble}
\affiliation{Gravitational Astrophysics Laboratory, NASA Goddard Space Flight Center, Greenbelt, MD 20771, USA}
\author{Julian~H.~Krolik} 
\affiliation{Physics and Astronomy Department, Johns Hopkins University, Baltimore, MD 21218, USA}
\author{Manuela~Campanelli} 
\affiliation{Center for Computational Relativity and Gravitation, Rochester Institute of Technology, Rochester, NY 14623, USA}
\author{Yosef~Zlochower} 
\affiliation{Center for Computational Relativity and Gravitation, Rochester Institute of Technology, Rochester, NY 14623, USA}
\author{Bruno~C.~Mundim} 
\affiliation{SciNet High Performance Computing Consortium, University of Toronto, Toronto, ON M5G 1M1, Canada}
\author{Hiroyuki~Nakano} 
\affiliation{Faculty of Law, Ryukoku University, Kyoto 612-8577, Japan}
\author{Miguel~Zilh\~ao}
\affiliation{Centro de Astrof\'\i sica e Gravita\c c\~ao - CENTRA, Departamento de F\'\i sica,Instituto Superior T\'ecnico - IST, Universidade de Lisboa - UL, Avenida Rovisco Pais 1, 1049-001, Portugal}

\begin{abstract}
Accreting supermassive binary black holes (SMBBHs) are potential
multi-messenger sources because they emit both gravitational wave and
electromagnetic (EM) radiation.  Past work has shown that their EM
output may be periodically modulated by an asymmetric density
distribution in the circumbinary disk, often called an ``overdensity''
or ``lump;" this modulation could possibly be used to identify a
source as a binary.  We explore the sensitivity of the overdensity to
SMBBH mass ratio and magnetic flux through the accretion disk.  We
find that the relative amplitude of the overdensity and its associated
EM periodic signal both degrade with diminishing mass ratio, vanishing
altogether somewhere between 1:2 and 1:5.  Greater magnetization also
weakens the lump and any modulation of the light output. We
  develop a model to describe how lump formation results from internal
  stress degrading faster in the lump region than it can be
  rejuvenated through accretion inflow, and predicts a threshold value
  in specific internal stress below which lump formation should occur
  and which all our lump-forming simulations satisfy.  Thus,
detection of such a modulation would provide a constraint on both
mass-ratio and magnetic flux piercing the accretion flow.
\end{abstract}


\section{Introduction}
Somewhere in the universe several pairs of supermassive black holes (SMBHs)
should merge every year, leaving behind a still more massive single
black hole at the centers of the galaxies where this occurs
\citep{Klein2016,Katz2020}.  These events are extremely challenging to
observe, but are of great interest because they are the most distant
gravitational wave sources we can hope to detect, and complementary
photon and gravitational wave data could provide uniquely powerful
diagnostics of these events \cite{Mangiagli2020,Baker2019,Kelley2019}.
In addition, the consequences of such mergers for galactic evolution
are profound, including strong correlations between the galaxies and
the (merged) central black holes.  Moreover, the physics of accretion
onto a binary is by no means limited to relativistic systems---it may
also be applied to protoplanetary disks in binary stellar systems
\cite{Keppler2020}.

A short time before merger, these systems are 
 in the post-Newtonian (PN) regime, in which the binary  loses energy and inspirals rapidly due to emission of gravitational radiation.  Because of their small separations (a fraction 
of a parsec) and the greater chance of a relatively rare SMBBH system being outside our local 
extragalactic neighborhood, spatially resolving the two BHs is unlikely. 
Hence, electromagnetic identification of SMBBH systems requires matching theoretical 
expectations to observed phenomena in their light curves, spectra, or polarization.

Accreting binaries whose mass-ratios $q \equiv M_2/M_1 \gtrsim 0.01$ generically exhibit a gap within a radius $\sim 2a$,
where $a$ is the binary's semi-major axis.   Matter travels across this gap in a pair of streams,
which convey the accreted mass to small accretion disks (``mini-disks") around each
member of the binary.   The binary can also break the axisymmetry of the circumbinary accretion disk \citep{MM08,Shi12,Noble12}, causing it to concentrate much of its inner rings' mass in a limited
range of azimuthal angles, a feature we refer to as the ``lump,'' following \citet{Shi12,Noble12}.
As a result, the accretion rate onto the binary is modulated
at a frequency $\simeq 0.2 \Omega_{\rm bin}$ (here $\Omega_{\rm bin}$ is the binary
orbital frequency).

Periodic modulation of the accretion rate can lead to a corresponding modulation
of the system luminosity
if the residence time of matter in the mini-disks is short compared to the modulation
period \citep{Bowen2018,dAscoli2018,Bowen2019}. 
Moreover, because the mass falling through the gap onto a black hole's
mini-disk is expected to shock against the outer edge of the mini-disk,
the hard X-rays radiated by that shocked gas should also be modulated
on the same time scale as the accretion rate \citep{Sesana2012,RoedigKM14}.

The heating rate directly associated with the lump is also modulated;
when $q=1$, the frequency is
twice the beat frequency between the binary's orbital frequency and the 
lump's orbital frequency \citep{Noble12}.  The lump is bright for the same reason
it exists.   As shown by \cite{Shi12}, it forms because some of the matter leaving the circumbinary
disk's inner edge gains enough angular momentum from the binary's gravitational torques
to travel back out to the circumbinary disk---rendezvousing there with the matter it left
shortly before.   The shock associated with its return drives heating within the lump.   The modulation frequency has this beat frequency because passing close to a black hole triggers inflow from the disk's inner edge.  In \cite{Noble12}, we put forth the 
idea that such a periodic signal could be used to identify SMBBHs photometrically and possibly 
constrain a SMBBH's mass ratio: the signal frequency would be less than twice the beat
frequency for unequal-mass binaries because the secondary BH predominantly interacts with the lump in those cases \citep{Noble12}.

Purported observations of periodic emission from AGN have been 
reported \citep{2015Natur.518...74G,2015ApJ...803L..16L,Charisi2016}.   In principle, they might originate from periodic phenomena similar to those just described.  However, the particular candidate identified in the first two of these studies did
not survive new data \citep{Liu2018,ZhuThrane2020}, and larger surveys have turned up marginal results
at best \citep{Liu2019,Chen2020,Liao2021}.   Because well-established periodic variation might yet be found, it is critical for us to understand the 
particular conditions from which such a signature arises, as well as the relation between
the period of such a modulation and the binary orbital period. 

Circumbinary black hole accretion simulations have been conducted in
numerous ways, and each way or method has demonstrated the development
of a lump or non-axisymmetric overdense feature under the right
conditions and measured the effect of mass ratio on accretion flow
properties.  Most of the simulations have been performed using 2-d
$\alpha$-model viscous hydrodynamics (2-d VH) and Newtonian gravity
using Eulerian grid-based codes
\citep{MM08,DOrazio13,Farris14a,Farris14b,DOrazio2016,Munoz2016,Miranda2017,Derdzinski2019,Munoz2019,Moody2019,Mosta2019,Duffell2020,Zrake2020,Munoz2020a,Munoz2020b,Tiede2020,Derdzinski2021}.
Others have used $\alpha$-viscosity hydrodynamics with Newtonian
gravity in 3-d, either using SPH
\citep{Ragusa2016,Ragusa2017,HeathNixon2020,Fontecilla2020,Ragusa2020}
or Eulerian grid-based methods \citep{Moody2019}.

Unfortunately, VH simulations use ad hoc internal stress
  models that poorly represent the expected angular momentum transport
  mechanism in real systems: internal magnetic stress.
  Further, the vast majority of VH work uses unrealistic isothermal
  equations of state or neglects the vertical extent of the system.
  These approximations have real consequences to predictions; for
  instance, VH simulations are never turbulent, are laminar, and
  exhibit relatively steady accretion flows, which is very different
  from the red-noise dominated variability always found in AGN light
  curves. 3-d MHD simulations, such as ours here, eliminate these
  approximations.  There have
  been a few MHD simulations using Newtonian gravity and an
isothermal equation of state
\citep{Shi12,Bankert2015,Shi15,Shi2016}, those using approximate
GR spacetimes to describe the binary's gravitational influence and our
thermodynamic model \citep{LopezArmengol2021}, those using high-order
PN gravity (such as ours here) \cite{Noble12,
  Zilhao2015,Bowen2018,Bowen2019}, and those using full numerical
relativity (NR) techniques (though not always evolving the spacetime
in order to hold the binary to fixed separations) and our
thermodynamic model \cite{Farris12,Gold14,Gold2014b,Paschalidis2021}
(see also the review by \cite{Gold2019}).

Only those using PN
  gravity and full NR have used more realistic thermodynamics and MHD.  The full NR simulations,
  however, are often too expensive to run for O(100) binary orbits that 
  precludes them from reaching relaxed conditions, demonstrating significant lump
  development, and covering a sufficiently vast temporal dynamic range
  needed for accurate variability analysis.  Our 3-d GRMHD simulations
  with PN gravity therefore falls into a particularly useful niche
  that allows one to include the most realistic physical assumptions
  for the lengths of time needed to explore the lump and variability
  in the circumbinary disk region.  

In this paper, we  begin 
to explore how the parameters characterizing circumbinary disks using 3-d GRMHD and PN gravity 
affect the nature of the signal.  Along the way, we will also attempt to further elucidate how the overdensity feature arises.
Our approach will be to use the simulation called RunSE in \cite{Noble12} as a benchmark,
contrasting it with new simulations having different parameters, but all sharing the identical
high-order PN spacetime---a binary black hole system with a non-evolving
circular orbit of separation $20M$.   These new simulations can be grouped into two sets.
 One is a survey of mass ratios: 1:1, 1:2, 1:5, and 1:10.  The second studies the effects of
 differing amounts of mass and magnetic flux  in the accretion disk.
 Our work here represents the first
  time anyone explores how mass ratio and magnetic field conditions affect the
  circumbinary disk with PN gravity and GRMHD techniques.  

The specifications of our simulations are given in
Section~\ref{sec:simulation-details}.  So that we can cleanly separate
the lump from smooth behavior, we begin our presentation of results
with a description of axisymmetric features
(Section~\ref{sec:axisymm-struct}).  In the following section
(Section~\ref{sec:non-axisymm-struct}), we report how distinctly lump
behavior depends on parameters.  With these results in hand, we are
able to discuss the dynamics controlling the growth of the lump
(Section~\ref{sec:orig-lump-accr}).  All the results are discussed
together and summarized in Section~\ref{sec:conclusions}.  Animations of the runs discussed in this paper may be found
online here: \url{https://youtube.com/playlist?list=PLNaEA0qwDBaeApzLr2oarVKiO3AFnklTV} .

\section{Simulation Details}
\label{sec:simulation-details}

In this section, we explain the methods we use to model our circumbinary 
accretion disk system and the parameters specifying its state.  The gravitational aspects of the simulation are described in 
Section~\ref{sec:spacetime}, while the aspects regarding the circumbinary magnetized gas 
are given in Section~\ref{sec:matter}. We discuss how we selected each run's parameters in Section~\ref{sec:run-details}.

We use geometrized units in which  $G=c=1$.  We will use Greek letters 
(e.g., $\mu,\,\nu,\,\lambda,\,\kappa$) 
to represent  spacetime indices $\left[0,\,1,\,2,\,3\right]$, and Roman letters 
(e.g., $i,\,j,\,k,\,l$) to represent spatial 
indices $\left[1,\,2,\,3\right]$.  

Although the unit of time in our calculations is $M$, the most important physical unit of time
is the binary orbit's dynamical time $\Omegabin^{-1}$.   This time scale depends on mass-ratio $q$, but only weakly. For instance, the relative difference between
the two most extreme orbital frequencies considered, $\Omegabin(q=1)$
and $\Omegabin(q=1/10)$, is $0.8\%$.  In this paper, when we present data all taken
from the same run, we will use the $q$-specific value for $\Omegabin$; when we 
discuss data in runs with different mass-ratio, we will ignore this
distinction.

\subsection{Binary Black-Hole Spacetime}
\label{sec:spacetime}

Because our investigation focuses on dynamics close to the SMBBH, we can safely ignore 
the gravitational influence of the disk's gas and assume that gravity is entirely
dictated by the binary.  As in our previous work \citep{Noble12,Zilhao2015}, we use a 2.5PN closed-form expression for the spacetime metric as described in \cite{Mundim:2013vca}.  Only the so-called ``Near Zone'' (NZ) metric is used because our numerical domain does not extend either close enough to the black holes or far enough from the binary for the other zones to be 
needed.   In the present work, unlike our previous, we will consider spacetimes in which the
mass ratio, $q=M_2/M_1 < 1$, where the  primary mass $M_1$ is always expected to be larger than that of the secondary BH ($M_2$), and  we set $M_1 + M_2 = M = 1$.  We therefore 
concentrate in this section on the spacetime's dependence on $q$.

It is easiest to write down the metric in PN harmonic coordinates, a Cartesian basis system.  Using the work of \cite{Blanchet:1998vx}, we find that the leading-order, non-trivial components
of the NZ metric are
\bea
g_{00}^{\rm NZ} &=&
-1
+ \frac{2 M_{1}}{r_{1}} 
+ \frac{2 M_{2}}{r_{2}} 
+ {\cal{O}}(v^4)
\,,
\cr
g_{0i}^{\rm NZ} &=&
-\frac{4 M_{1}}{r_{1}} v_{1}^{i} 
-\frac{4 M_{2}}{r_{2}} v_{2}^{i} 
+ {\cal{O}}(v^5)
\,,
\cr
g_{ij}^{\rm NZ} &=& 
\delta_{ij} 
+ \frac{2 M_{1}}{r_{1}} \delta_{ij}
+ \frac{2 M_{2}}{r_{2}} \delta_{ij}
+ {\cal{O}}(v^4)
\,,
\label{eq:NZmetric}
\eea
where $v_{N}^{i}$ denotes the coordinate velocity of the $N^\mathrm{th}$ BH, and
$r_N=|\vec{x}-\vec{y}_N|$, i.e., computed from the Cartesian PN harmonic
coordinates.   Even though the metric is most simply
  represented in this Cartesian basis, in our simulation it is transformed
  to a spherical basis for use in our spherical coordinate system.

 The Near-Zone metric is valid only at distances more than $10 M_N$ from the $N^\mathrm{th}$ black hole
 \citep{Yunes:2006iw,Yunes:2005nn,JohnsonMcDaniel:2009dq}.
 This constraint means that $\rin$, the inner radial coordinate of our numerical domain, is bound from below to ensure the the metric's validity.  It can be estimated through the following argument.
The positions of the two BHs relative to the center-of-mass are 
\bea
&& \vec{y}_{1} = \frac{M_{2}}{M} \vec{a} + {\cal{O}}(v^{4})
= \frac{q}{1+q} \vec{a} + {\cal{O}}(v^{4}) \,,
\cr
&& \vec{y}_{2} = - \frac{M_{1}}{M} \vec{a} + {\cal{O}}(v^{4}) 
= - \frac{1}{1+q} \vec{a} + {\cal{O}}(v^{4}) \,,
\eea
where $\vec{a}$ denotes the separation vector from the secondary BH to the primary BH,
and we have ignored the 2PN order correction.   The constraint on $\rin$ is then
\bea
\rin\gtrsim a\frac{1 + q/2}{1+q} \,.
\label{eq:rin_vs_a}
\eea
The ratio $\rin/a$  increases from 0.75 for $q=1$ to 0.95 for $q=0.1$.

The NZ metric itself also depends on $q$:
\begin{align}
g_{00}^{\rm NZ} =&
-1
+ \frac{2}{(1+q)r_{1}} 
+ \frac{2q}{(1+q)r_{2}} 
\cr
=& 
-1
+ \frac{2}{r}
+{\frac {q{a}^{2} }{ \left( 1+q \right) ^{2}{r}^{3}}} [ 3\, \left( \vec{n} \cdot \hat{a} \right)^{2}-1 ]
\,,
\cr
g_{0i}^{\rm NZ} =&
\frac{4 q}{(1+q)^2 \sqrt{a}}
\left(
-\frac{1}{r_{1}}
+\frac{1}{r_{2}} \right) (\vec{\lambda})^i
\cr
=&
\frac{4 q}{(1+q)^2 \sqrt{a}}
\left(
-{\frac {a }{{r}^{2}}}\, \vec{n} \cdot \hat{a} \right.
\cr & \left.
+{\frac {{a}^{2} \left( 1-q \right) }
{ 2 \left( 1+q \right) {r}^{3}}}[ 3\, \left( \vec{n} \cdot \hat{a} \right) ^{2}-1 ]  \right) (\vec{\lambda})^i
\,,
\cr
g_{ij}^{\rm NZ} =& 
\left( 1 
+ \frac{2}{(1+q)r_{1}}
+ \frac{2q}{(1+q)r_{2}} \right) \delta_{ij}
\cr
=& 
\left( 1
+ \frac{2}{r}
+{\frac {q{a}^{2} }{ \left( 1+q \right) ^{2}{r}^{3}}} [ 3\, \left( \vec{n} \cdot \hat{a} \right) ^{2}-1 ] \right) \delta_{ij}
\,,
\end{align}
where $\vec{n}=\vec{r}/r$, $\hat{a}=\vec{a}/a$, $\vec\lambda$ is the unit vector tangent
to the black holes' orbit, and we have included only the non-trivial leading-order PN terms.
For each component, the second line shows a series expansion with respect to $a/r$
up to $O((a/r)^2)$ beyond the leading-order nontrivial contribution.

The quantity $q/(1+q)^2$, which appears in every metric component, 
decreases monotonically as $q$ falls from $1$ to $0$.  The quadrupolar component in
$g_{00}^{\rm NZ}$ and $g_{ij}^{\rm NZ}$, which has an angular dependence
$[3\, \left( \vec{n} \cdot \hat{a} \right) ^{2}-1]$, therefore diminishes for smaller mass ratio.
On the other hand, the quadrupolar component in $g_{0i}^{\rm NZ}$,
which is a higher-order PN contribution to the equations of motion,
has a maximum for $q \approx 0.27$.  For further details about the range of validity for the NZ metric, see Appendix~\ref{app:NZValidity}.

\subsection{Matter}
\label{sec:matter}

We use the flux-conservative code \harm to evolve the GRMHD equations on the dynamical spacetime \citep{GMT03,Noble06,Noble09,Noble12}.  It uses piecewise parabolic reconstruction of the primitive variables at each cell interface, a  Lax-Friedrichs-type flux, and a 3-d version of the 2-d FluxCT constrained transport scheme originally described in \cite{Toth00}.   The stationary spacetime version, the same MHD methods as we use here, was described in \cite{Noble10}, while the modifications for 
handling dynamic spacetimes were explained in \cite{Noble12}.  We refer the reader to these papers for further details on our numerical algorithms, and only briefly describe the equations of motion (EOM) for purposes of definition. 

The EOM dictating magnetized gas's evolution are the Euler-Lagrange-Maxwell
equations on a curved background spacetime with metric $g_{\mu \nu}$
The entire set may be written in the manifestly conservative form
\beq{ 
\del_t \bU\left(\prim\right) = -\del_i
  \bF^i\left(\prim\right) + \mathbf{S}\left(\prim\right) \, ,
\label{conservative-eq}
}
where $\bU$ is a vector of ``conserved'' variables, $\bF^i$ is a vector of fluxes, 
and $\mathbf{S}$ is a vector of source terms:
\beq{
\bU\left(\prim\right) = \sqrt{-g} \left[ \rho u^t ,\, {T^t}_t 
+ \rho u^t ,\, {T^t}_j ,\, B^k  \right]^T
\label{cons-U}
}
\beq{
\bF^i\left(\prim\right) = \sqrt{-g} \left[ \rho u^i ,\, {T^i}_t + \rho u^i ,\, {T^i}_j ,\, 
\left(b^i u^k - b^k u^i\right) \right]^T
\label{cons-flux}
}
\beq{
\mathbf{S}\left(\prim\right) = \sqrt{-g} 
\left[ 0 ,\, 
{T^\kappa}_\lambda {\Gamma^\lambda}_{t \kappa} - \mathcal{F}_t ,\, 
{T^\kappa}_\lambda {\Gamma^\lambda}_{j \kappa} - \mathcal{F}_j ,\, 
0 \right]^T \, .
\label{cons-source}
}
Here $B^i = \dF^{it}/\sqrt{4\pi}$ is the magnetic field, 
$\dF^{\mu  \nu}$ is the Maxwell tensor, 
$u^\mu$ is the fluid's $4$-velocity, and
$b^\mu = \frac{1}{u^t} \left({\delta^\mu}_{\nu} + u^\mu u_\nu\right)
B^\nu$ is the magnetic $4$-vector (which is the magnetic field when projected into
the fluid's co-moving frame).
In addition, $W = u^t / \sqrt{-g^{tt}}$ is the fluid's Lorentz function, 
$\mathcal{L}$ is the fluid-frame bolometric energy loss rate, and
$\mathcal{F}_\mu = \mathcal{L} u_\mu$ is the radiative flux 4-vector.
Lastly, $g$ is the determinant of the metric, and
${\Gamma^\lambda}_{\mu \kappa}$ is the metric's affine connection.  The
stress-energy tensor $T_{\mu \nu}$ is defined as
\beq{
T_{\mu \nu} = \left( \rho h + 2 p_{m} \right) u_\mu u_\nu   + \left( p + p_{m}\right) g_{\mu \nu} - b_\mu b_\nu 
\quad , 
\label{mhd-stress-tensor}
}
 where $p_m = b^\mu b_\mu / 2$ is the magnetic pressure, $p$ is the
gas pressure, $\rho$ is the rest-mass density, $h = 1 + \epsilon +
p/\rho$ is the specific enthalpy, and $\epsilon$ is the specific
internal energy.

Often, we look at reduced properties of the accretion flow and examine
spherical shell-averaged  or shell-integrated quantities.  Shell averages are made using 
\beq{ \shellavg{X} \equiv
  \frac{\int X \, \sqrt{-g} \, d\theta d\phi}{\int \sqrt{-g} \,
    d\theta d\phi} \, ,
\label{shell-average-def}
}
while shell-integrated quantities are defined by
\beq{
\shellint{X} \equiv \int X \sqrt{-g} \,  d\theta \, d\phi \ = \shellavg{X}  \int \sqrt{-g} \, d\theta \, d\phi 
\,. \label{shell-integral}
}
Mass-weighted shell-averaged quantities are denoted as 
\beq{
\massavg{X} \equiv \frac{ \shellint{X \rho} }{ \shellint{ \rho }}
\,. \label{density-shell-averaged}
}
We also sometimes smooth time-dependent quantities to highlight trends over longer time scales, and these are designated by overbars:
\beq{
\timesmooth{X}{t, \Delta t} = \frac{1}{\Delta t} \int^{t + \Delta t / 2}_{t - \Delta t / 2} X(t^\prime) dt^\prime
\quad . \label{tsmooth-def}
}

\subsection{Run Details}
\label{sec:run-details}

The parameters defining the runs' configurations are stated in
Table~\ref{tab:init-parameters}.  Measured quantities resulting from
the runs are given in Table~\ref{tab:run-characteristics}.

Our goal is to define how ``lump" behavior (and a few other aspects of circumbinary
disk dynamics) depend on mass-ratio and
magnetic content of the disk.   To that end, we made every new simulation
as similar as possible to our fiducial simulation, \runse of \cite{Noble12}.
However, technical considerations demanded small departures in certain instances.

The binary was held at a fixed separation of $20M$ in all cases, like
\runse, so that we can explore the development of the lump and the
quality of the periodic EM signal from it.  Also like \runse,
in all runs, the binary orbit was circular, and the black holes were non-spinning.

We endeavored to make the problem spacetime volume very nearly the same as in \runse ,
extending radially from an inner excision radius of $0.75a$ to an outer boundary radius of
$13a$.  However, in the three runs with $q < 1$ the inner excision radius was moved outward 
to $a$ in order to accommodate the orbit of the secondary black hole, and in two runs 
the outer boundary was pushed out to $50a$ in order to contain a larger disk.
The larger value of $\rin$ is within the limit---$\rin/a=1.1$---at which \cite{Shi12} 
began to see it significantly alter the structure of the circumbinary disk's inner edge.
In every case, we used the same angular excision around the polar axis as in \runse,
and, as is necessary for binary dynamics, all runs covered a full $2\pi$ in azimuth.
All the new runs had durations at least as long as the original \runse
run, from $1.25\times$  to more than $2\times$ that of \runse.

The angular grid in every run was the same as
in \runse, with $N_\theta = 160$~cells in polar angle and $N_\phi =
400$~cells in azimuthal angle.   The disks always satisfied the MRI
quality conditions of \cite{hgk11} well, so there was no need to
increase the number of points.   The ratio $\Delta r/r$ was the same
in every run, and the number of cells $N_r = 300$ was also the same in
all but two.  The outer radius of our numerical grid, $\rout$, was
chosen to lie beyond the extent of the initial distribution of gas.
With $\Delta r/r$ fixed, \medrun and \lrgrun
required more radial zones to reach their extended outer boundaries.

The initial gas distribution of each run was an equilibrium state with respect to the
time-average of the run's spacetime, in which the gas was supported against gravity by
pressure gradients and rotation.   Appendix~A of
\cite{Noble12} provides the details of how the hydrostationary solutions
were found.  The parameters that constrain the solution are: the radius
of the inner edge of the disk $\rdiskin$, the radius of the pressure maximum
$\rpmax$, and the initial aspect ratio ($H/r$) of the disk at the pressure maximum.
All of these were identical to the values of \runse, except in the
two runs studying larger disks, where $\rpmax$ was larger by 20--25\% and
$\rdiskout$ was larger by a factor 2--3.  In all cases, $H/r = 0.1$.
The disks were initially isentropic,
with entropy $K = p/\rho^\Gamma = 10^{-2}$ and adiabatic index $\Gamma=5/3$.
All simulations used an ideal-gas equation of state with $p=\left(\Gamma-1\right) u$
for internal energy density $u$.

In every run, the initial magnetic field was given a poloidal
distribution in the same way as in \cite{Noble12}. The magnetic
field amplitude was normalized so that the ratio of the
volume-integrated magnetic and gas pressure, an approximation to the
volume-averaged plasma $\beta=p/p_m$, was $100$.

So that we can tally the amount of energy dissipated during evolution, all runs cool to the same target entropy, which we choose to be the initial entropy, uniform throughout the flow.   We parameterize this entropy by a proxy $K \equiv p/\rho^\gamma$.   Its value in the initial state $K_0=10^{-2}$.   A fluid element 
is cooled if its entropy is above $K_0$, and neither cooled nor heated if $K < K_0$.  Writing $K=K_0 + \Delta K$, the cooling rate is
\beq{
\mathcal{L} = \frac{\rho \epsilon}{T_\mathrm{cool}} \left( \frac{\Delta K}{K_0} + \left|\frac{\Delta K}{K_0}\right| \right)^{1/2}  \quad .
\label{cooling-function}
}
Note that \cite{Noble12} erroneously omitted the exponent of $1/2$ in the paper, though they used it in the simulations reported there. 
The cooling time is  $T_\mathrm{cool} =  2 \pi \left(r/M\right)^{3/2}$,
the period of a circular equatorial orbit at radius $r$ without any quadrupolar contributions to the potential.

The cooling rate is recorded 
as 3-d data for the duration of each simulation, so that it may serve as a proxy for 
the gas's bolometric emissivity.

\begin{table}
  \hspace{-1.8cm}
  \begin{tabular}{|lccccccc|}
\hline
\textbf{Name} &$q$  &$N_r$  &$\rin$  &$\rout$  &$\rdiskin$  &$\rpmax$  &$\rdiskout$ \\ 
\hline 
\hline 
\origrun        &$1  $  &$300$   &$0.75$ &$13$  &$3$ &$5$   &$11.7$  \\
\qtwo           &$0.5$  &$300$   &$1$    &$13$  &$3$ &$5$   &$11.7$  \\
\qfive          &$0.2$  &$300$   &$1$    &$13$  &$3$ &$5$   &$11.7$  \\
\qten           &$0.1$  &$300$   &$1$    &$13$  &$3$ &$5$   &$11.7$  \\
\medrun         &$1  $  &$400$   &$0.75$ &$50$  &$3$ &$6$   &$23.4$  \\
\lrgrun         &$1  $  &$420$   &$0.75$ &$50$  &$3$ &$6.5$ &$39.1$  \\
\injectrun      &$1  $  &$300$   &$0.75$ &$13$  &$3$ &$5$   &$11.7$  \\
\hline 
\end{tabular}
\caption{Parameters determining the simulations of the \massratioruns
  and \magfluxruns.  All radii are given in units of the binary
  separation, $a=20\mbh$.}
\label{tab:init-parameters}
\end{table}

\begin{table}
  \hspace{-1.4cm}
\begin{tabular}{|lcccc|}
\hline
\textbf{Name} &\textbf{$\Sigma_0$} &\textbf{$\omegalump [\Omegabin]$} &$\Tlump [10^3 \mbh]$  &$\tend [10^3 \mbh]$ \\ 
\hline 
\hline 
\origrun        &$0.096$ &$0.27 \pm 0.05$ &$52$  &$76$ \\
\qtwo           &$0.083$ &$0.26 \pm 0.05$ &$77$  &$107$ \\
\qfive          &$0.082$ &---             &---   &$95$ \\
\qten           &$0.082$ &---             &---   &$97$ \\
\medrun         &$0.085$ &$0.25 \pm 0.09$ &$91$  &$143$ \\
\lrgrun         &$0.087$ &$0.26 \pm 0.12$ &$133$  &$158$ \\
\injectrun      &$0.096$ &$0.25 \pm 0.26$ &$113$  &$126$ \\
\hline 
\end{tabular}
\caption{ Measured characteristics of the simulations of the \massratioruns
  and \magfluxruns.  The circumbinary disk's initial peak surface
  density, $\Sigma_0$, is in units of code units for density times $\mbh^{2}$.  }
\label{tab:run-characteristics}
\end{table}


With all these quantities fixed, we performed two series of parameter-exploration runs, one varying the black hole mass ratio, the other varying the initial magnetic
flux given the disk. 

\subsubsection{\massratiorunscap}
\label{sec:mass-ratio-survey-1}
Real supermassive black hole binaries can have a variety of
mass ratios. To measure the effect the mass ratio has on the
circumbinary flow, we performed a series of simulations labeled \qone,
\qtwo, \qfive, and \qten, having mass ratios 
$q=\mbh_2/\mbh_1 = \left\{\, 1, \, 1/2, \, 1/5, \, 1/10 \, \right\}$, 
respectively, but all other physical parameters the same.
This set was chosen in the hope of
covering the whole range relevant to gas accretion.  When $q\rightarrow0$, 
the secondary black hole acts only as a mild perturber, producing
little effect on the circumbinary flow.  In fact, well before it reaches that
limit, as we will show later in the
paper, small $q$ leads to weaker overdensities: even at $q=1/5$, the lump
amplitude is quite small if detectable at all.
Simulations \qone, \qtwo, \qfive, and \qten ran until approximately
$76\mathrm{kM}$, 
$107\mathrm{kM}$, 
$95\mathrm{kM}$, and
$97\mathrm{kM}$, respectively.   

\subsubsection{\magfluxrunscap}
\label{sec:magnetic-flux-series}

The idea that parts of an accretion disk can have regions of low
angular momentum transfer, like the lump, is not a new one.  For
instance, in protoplanetary disks, the midplane of the accretion disk
may be so shielded from cosmic radiation and its central source that
it may be too cold to be adequately ionized and magnetized.  Because
accretion manifests from angular momentum transport mediated by
magnetic stresses, such ``dead-zones'' will be uncoupled from the rest
of the disk \cite{1996ApJ...457..355G,Gole2016}.  The transition from
an actively accreting region to inactivity leads to a build up of
matter, similar to the development of our overdensity.  Just as a
protoplanetary dead-zone may be revived by a local heating event
thereby turning active again, so may our ``dead'' overdensity be
eroded away if given an injection of additional magnetic flux.  We
present new runs here designed to see whether added magnetic flux may
reignite activity in the dead zone or overdensity.

The control run for this series is also \origrun from \cite{Noble12}.
From this run, we learned that
the growth in amplitude of the overdensity feature was coincident
in space and time with a decline in  how well the simulation
can resolve the MHD turbulence, as defined by the quality factor $Q^i \equiv v^i_{A}/(\Delta x^i \Omega_K)$, where $v^i_A$ is the Alfven speed associated with the $i$-th magnetic field component, $\Omega_K$ is the local circular orbit frequency, and $\Delta x^i$ is the cell-size in the $i$-th direction \citep{Noble10,hgk11}.
Here $v_{Ai}$ is the Alfven speed for the magnetic field component in the $i^\mathrm{th}$-direction and
$\Delta x^i$ is the cell-size in that dimension. Two mechanisms, logical converses of each other, may explain this effect:
\begin{enumerate}
  \item  A decline in local MHD stress per unit mass (signaled by decreasing $Q$)
fosters the growth of the overdensity.
\item The lump's increasing density decreases $v_A$, degrading
the effective resolution of the simulation; this numerical effect then retards, or even eliminates, magnetic field growth.
\end{enumerate}
It is also entirely possible that both act, reinforcing one another.

The first possibility implies the correlation is physical and our results are
potentially predictive. 
The second possibility implies that our simulations have little
predictive power since we cannot say whether higher resolution (i.e.,
what nature uses) would yield an overdense feature.

With three additional runs, we aim to test whether either of
these mechanisms operates.  In \injectrun, we ask whether a late-time
strengthening of the magnetic field can, by restoring resolution quality,
sustain magnetic stress despite increasing gas density in the lump.
In \medrun and \lrgrun, we increase the total magnetic flux available to
the disk to test whether stronger field retards lump growth.

For \injectrun, at $t=5\times 10^4\mbh$ we added to the existing magnetic
field additional poloidal field whose geometry matches that of the
initial field, i.e.,
\beq{ 
  B^i \ = \ B^i_{O2} + f
  \, B^i_{O1} \frac{\sqrt{-g}_{O1}}{\sqrt{-g}_{O2}}
  \quad , \label{injected-mag-field} 
}
where the subscript ``O1'' (``O2'') means the quantity comes from
\origrun at $t=0\mbh$ ($t=5\times10^4\mbh$). This procedure automatically
preserves the solenoidal character of the field, while also minimizing
significant transient behavior. Adding a poloidal field is also desirable
because, for equal field intensity, poloidal field leads to more rapid
MRI growth than toroidal field \cite{HK02,BHK08}.
We set the constant factor $f=2$ to make the field dynamically significant
after an orbital time scale. The ratio of $\sqrt{-g}$ at different times is
necessary in our case because $\sqrt{-g}$ is time-dependent, and the
determinant is included in the covariant form of the magnetic field's
constraint equation:
\beq{
  \nabla_\mu B^\mu = \frac{1}{\sqrt{-g}} \partial_i \left( \sqrt{-g}
  \, B^i \right) = 0 \quad .  
} 

Because the magnetic field in the late-time snaphot of \origrun is
turbulent, the magnetic field has a large dynamic range; consequently, the
added ordered magnetic field may change the field locally by an amount
 $\mathcal{O}(1)$.  However, the total magnetic field energy
added to the system through this procedure is less than $7\%$ 
of the existing energy.
Once the magnetic field is added, the disk is allowed to evolve for an
additional $7.5 \times 10^4M$ in time.

In \medrun and \lrgrun we increased the initial reservoir of magnetic
flux available by increasing the size of the hydrostationary torus
that encompasses the initial poloidal magnetic field distribution.
The initial extent of the magnetic field distribution and its
integrated flux content all scale with the size of the disk.   Because
these disks also have a larger mass reservoir, we expect them to
sustain longer periods of accretion than what we observed with
the smaller disk (Section~\ref{sec:mass-accretion-rate}).
This longer run time also helps to eliminate a concern that the overdensity
develops because diminishing mass accretion at late times
also leads to diminishing magnetic flux delivery, and therefore might permit a longer-lived
lump.  \medrun (\lrgrun) included $37\%$ ($78\%$) more magnetic flux and $40\%$ ($72\%$) more mass than \origrun.
  

\section{Axisymmetric Structure}
\label{sec:axisymm-struct}

In order to justify our reliance on time-averages and make comparisons
to steady-state disk theory, we need to evaluate how well our
simulations have reached an equilibrium with respect to the accretion
of mass.  This is important to observables (e.g., electromagnetic
luminosity) because the emissivity is proportional to the local
rest-mass density \cite{Noble10,Noble12,dAscoli2018}.
Further, if the system fluctuates with $\mathcal{O}(1)$
fluctuations on time scales comparable to our simulation's
duration, then our results have little predictive power.  To assess
the degree to which a simulation has entered a state of
mass-inflow equilibrium we will measure: the accretion
history, the time evolution of surface density, and the history of the
integrated mass at sample radii. These are all most efficiently
evaluated using poloidal- and azimuthally- integrated quantities, the
focus of this section.

\subsection{Mass Accretion Rate}
\label{sec:mass-accretion-rate}

\begin{figure}[htb]
\centerline{
\includegraphics[width=\columnwidth]{\plotdir/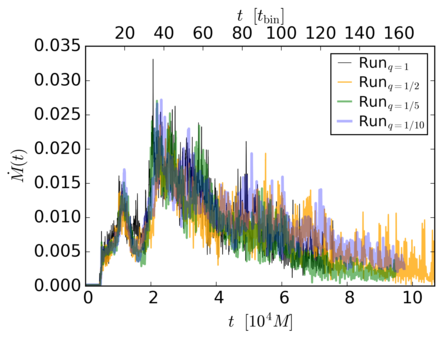}
}
\caption{Mass accreted $r=a$ as a function of time, $\dot{M}(r=a,t)$, for the \massratioruns.
}
\label{fig:mdot-vs-t-mass-ratios}
\end{figure}

\begin{figure}[htb]
\centerline{
\includegraphics[width=\columnwidth]{\plotdir/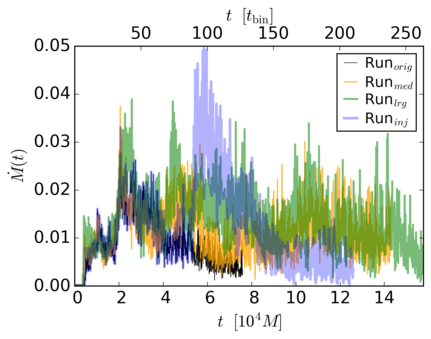}
}
\caption{Mass accreted $r=a$ as a function of time, $\dot{M}(r=a,t)$, for the \magfluxruns.
}
\label{fig:mdot-vs-t-states}
\end{figure}

The accretion rate history through the spherical surface at $r=a$ is
shown in
Figures~\ref{fig:mdot-vs-t-mass-ratios}~and~\ref{fig:mdot-vs-t-states}.
All runs show a gradual decline of the accretion rate after a period
of rapid accretion early on.  The peak is due to the burst of inflow
that occurs after the linear growth of the MRI saturates.  
The decline is slow and has only minor fluctuations, allowing us to scale out this slowly
varying secular trend when necessary.

Mass ratio does not appear to have a large effect on the accretion
history compared to the accretion rates' variability at any one
instant. The similarity between the curves highlights the fact that
the long time-scale trend is primarily dictated by the initial
conditions of the torus.

The accretion rate's dependence on the  initial physical
state of the torus is emphasized in
the \magfluxruns.  Although the accretion rates of all the runs in this
group are very similar within the first $3\times10^4 \mbh$ of time,
at later times they develop larger fluctuations and no longer mimic
one another so closely.   Nonetheless, all but \injectrun may be fairly
described as having a late-time accretion rate that fluctuates within the
range $\approx 0.01 - 0.02$, a rate several times greater than the long-term
accretion rate for the runs in the mass-ratio series.
The larger mass reservoirs of \medrun and \lrgrun explain the greater
sustained rates of accretion.  

The addition of ordered magnetic field at  the run's start, $t=5\times10^4 \mbh$,
makes \injectrun different from the others.  The burst
of accretion seen in \injectrun at $t \approx 5.5 \times 10^4\mbh$ is triggered
by the added ordered magnetic field; the delay is the time required for MRI
growth to amplify the MHD turbulence to the saturation level associated
with the larger magnetic flux.

\begin{figure}[htb]
\centerline{
\includegraphics[width=\columnwidth]{\plotdir/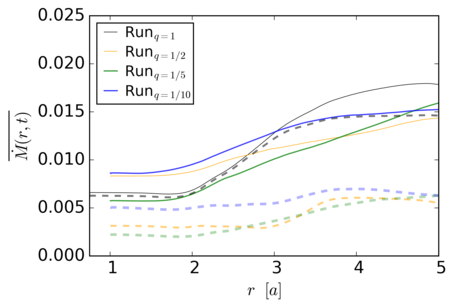}
}
\caption{Time average of the mass accretion rate as a function of $r$ for the \massratioruns.
  The time interval used for the averages is $40,000 < t/M < 76,000$ (solid curves),
  the secularly-evolving period of \qone, and the last $3\times10^4\mbh$ of each run (dashed curves).}
\label{fig:mdot-vs-r-mass-ratios}
\end{figure}

A key question to answer in regard to circumbinary accretion is how
much of the mass accretion rate in the outer disk can
penetrate into the domain of the binary despite the binary's gravitational torque.
This quantity, most often called the ``leakage fraction", gauges
the degree to which the overall system is in inflow equilibrium.  Most
of the effort on this topic hitherto used hydrodynamics simulations
employing a phenomenological ``$\alpha$" viscosity to transfer
angular momentum in 2-d Eulerian codes
\citep{MM08,DOrazio13,Farris14a,Munoz2016,Miranda2017,Tiede2020,Duffell2020}
or in 3-d Smoothed Particle Hydrodynamics (SPH) codes \citep{Ragusa2016,HeathNixon2020}; 
much less attention has been given to inflow dynamics resulting from genuine MHD stresses
\citep{Shi12,Noble12,Zilhao2015,Gold14,Shi15}.
The consensus from both
hydrodynamic and MHD work for any explored mass ratio is that the ``leakage fraction"
is essentially unity;
in other words, the system reaches inflow equilibrium.\footnote{Only \citet{Ragusa2016} and \citet{HeathNixon2020} dissent from this view.}  The ratio between the mass accretion rate
and the peak surface density near the inner edge may, however, depend on the disk aspect ratio
\citep{Tiede2020}.

\begin{figure}[htb]
\centerline{
\includegraphics[width=\columnwidth]{\plotdir/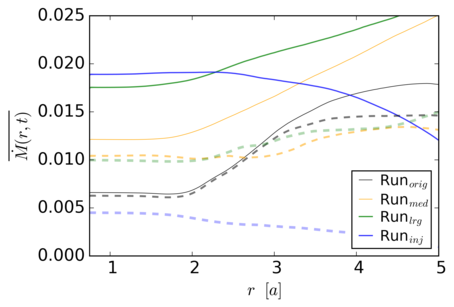}
}
\caption{ Time average of the mass accretion rate as a function or
  $r$ for the \magfluxruns.  The time interval used for the averages is $40,000 < t/M <
  76,000$ (solid curves), the secularly-evolving period of \qone, and
  the last $3\times10^4\mbh$ of each run (dashed curves).}
\label{fig:mdot-vs-r-states}
\end{figure}

Figures~\ref{fig:mdot-vs-r-mass-ratios} and \ref{fig:mdot-vs-r-states} portray our
results on this question.   For all the new runs in the mass-ratio series, the extended
duration of our simulations led to a significant improvement in the quality of
inflow equilibrium.   Cases with smaller $q$ generally have  smaller departures
from inflow equilibrium, but by the end of all the new simulations, the accretion
rate became reasonably close to constant as a function of radius out to $\simeq 5a$.

The time-averaged accretion rate as a function of radius
is displayed for the \magfluxruns in
Figure~\ref{fig:mdot-vs-r-states}. Again, we see the same
flattening of $\mdotvsr{r}$ over time in these runs.  Both \medrun and
\lrgrun asymptote to similar profiles at late times. 
Because they were both run to longer times,
this fact supports the notion that eventually $\mdotvsr{r}$ asymptotes to
a flat profile in all cases.

The accretion rate profile from \injectrun contrasts
strongly with all the others, curving downward with radius.  As we will see in further analysis,
the magnetic field perturbation of \injectrun leads to a sudden
accretion episode that drains a majority of the available mass in the
domain. After this happens, the torus in this run no longer has a mass
reservoir able to sustain mass-inflow equilibrium.

\subsection{Enclosed Mass}
\label{sec:enclosed-mass}

\begin{figure}[htb]
\centerline{
\includegraphics[width=\columnwidth]{\plotdir/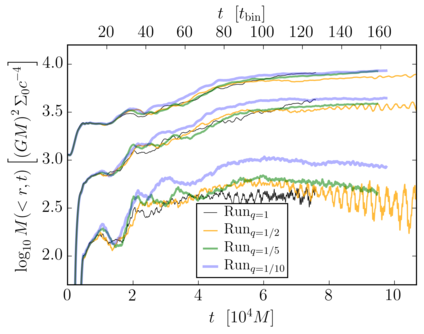}
}
\caption{Mass enclosed within a set of radii as a function of time for the \massratioruns.
  The sample radii are at $r/a = 2, 3, 4$ from bottom to top.
}
\label{fig:mass-enclosed-mass-ratios}
\end{figure}

The next means by which we evaluate mass inflow equilibrium is through
the mass enclosed within a given radius versus time, $M(<r,t)$:
\beq{
  M(<r,t) \equiv \int^r_{\rin} dr^\prime \int d\Omega \sqrt{-g} \, \rho  \quad .
  \label{enclosed-mass}
}
In Figure~\ref{fig:mass-enclosed-mass-ratios}, we plot these trends at sample
radii $r/a = 2, 3, 4$ for each mass ratio run. The $M(<r,t)$ curves
for disks in perfect equilibrium at all radii should all be flat.
Therefore, deviations from constancy indicate departures from inflow equilibrium.
As we found previously for \qone, the mass within $r=2a$ hardly
changes throughout the post-transient, secularly evolving period (i.e., $t\simeq
5\times10^4\mathrm{M}$); this is also true to within 10\% for all the other mass ratios.
We also find that all the
$q<1$ runs exhibit flatter $M(<r,t)$ trends in time at the largest enclosed radii,
with all becoming nearly flat by $t=\tend$; we provide the values of $\tend$ for all runs in Table~\ref{tab:run-characteristics}.
This implies that each simulation in the
\massratioruns is in approximate mass inflow equilibrium out to these radii.
Fluctuations in enclosed mass for all runs are strongest at the smallest radii and
grow weaker with decreasing mass ratio, which is expected since the
magnitude of the binary's gravitational torque decreases with distance
and mass ratio.

The small decline of $M(<r,t)$ from peak to $\tend$ decreases in
magnitude as the mass ratio decreases, with only $\sim3\%$ change for
\qten.  We also see that \qten has significantly more mass enclosed at
radii $r/a=1.5,2$ for all times. This implies that there is a more
massive distribution of steady gas within these smaller annuli. As we
will see, this is consistent with the fact \qten has a fuller
``cavity'' region.

These measures taken together suggest that all the mass ratio simulations exhibit
a (weakly) secularly-evolving steady-state of mass flow within $2<r/a<4$ for
$t\gtrsim5\times10^4\mbh$.

\begin{figure}[htb]
\centerline{
\includegraphics[width=\columnwidth]{\plotdir/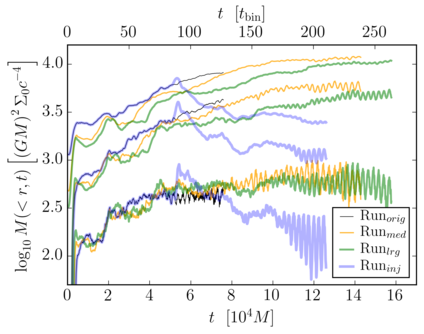}
}
\caption{Mass enclosed within a set of radii as a function of time for the \magfluxruns.
  The sample radii are at $r/a = 2, 3, 4$ from bottom to top.
  }
\label{fig:mass-enclosed-states}
\end{figure}

For the \magfluxruns, the trends in enclosed mass are shown in
Figure\ref{fig:mass-enclosed-states}.  For the larger radii,
$r/a=3,4$, the mass-enclosed curves for \origrun rise steadily all the
way to $t=t_{\rm end}$; the curves for \medrun and \lrgrun follow closely
that of \origrun until its end, but plateau, and to the same level, at
$t\simeq 1.2 \times 10^5\mbh$.
In fact, $M(<r,t)$ for \medrun and \lrgrun are also nearly identical at late times at $r/a=1.5,2$
suggesting that these runs have likewise achieved steady-state at
these smaller radii.  Because their initial distributions of mass and magnetic flux
are quite different, this late-time matching suggests that
their mass inflow equilibrium is generic for this
thermodynamic model. 

The \injectrun $M(<r,t)$ trends are significantly different from those
of any other run, however. After the time of injection, we see a
dramatic increase in mass followed by a rapid decline for all probed
radii $r<4a$.  In contrast, all runs but \injectrun had growing $M(<r,t)$ at
$r/a=3,4$.  Apparently, the magnetic field injection resulted in a
sudden redistribution of mass inward that left the entire circumbinary disk with
significantly less mass than in the other runs.  

\subsection{Surface Density Evolution}
\label{sec:surf-dens-evol-mass-ratios}

One of the key differences between circumbinary disks and disks around
a single black hole is the very low
surface density gap
carved out of the accretion flow when there is a binary at its center.
The surface density is conventionally defined as the rest-mass density integrated along
its ``vertical height.''  For our
relatively thin disks aligned with the binary's angular momentum, the
vertical integral is approximated well by an integral along the
poloidal direction:
\beq{ \Sigma(r,\phi,t) = 
  \frac{ \int d\theta \sqrt{-g} \, \rho }{\left.\sqrt{g_{\phi
        \phi}}\right|_{\theta=\pi/2}}
  \quad . 
  \label{surface-density-vs-r-phi}
}
Often it is useful to examine azimuthal averages of the surface density, which we calculate as:
\beq{
  \Sigma(r,t) = 
  \frac{ \left\{ \rho \right\} }{\int d\phi \left.\sqrt{g_{\phi \phi}}\right|_{\theta=\pi/2}}  \quad . 
  \label{surface-density-vs-r}
}
It can be convenient to measure the surface density in units of the initial peak surface density, $\Sigma_0$; the values of $\Sigma_0$ for all runs are given in Table~\ref{tab:run-characteristics}.

\begin{figure*}[htb]
\centerline{
\includegraphics[width=2\columnwidth]{\plotdir/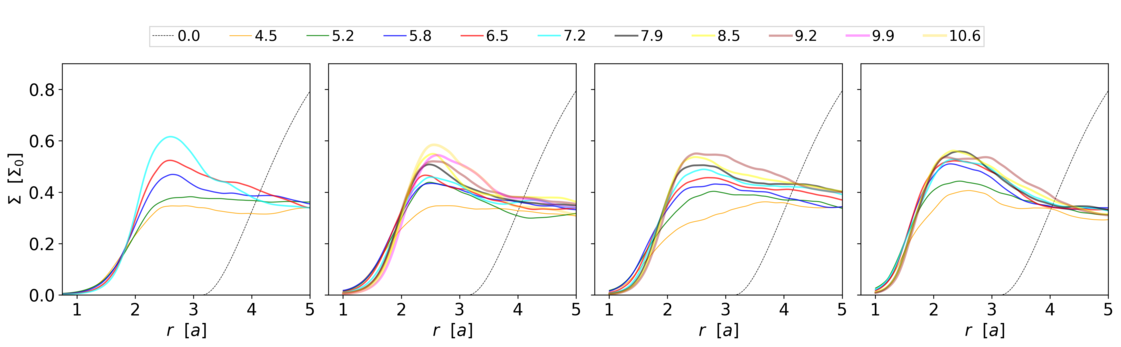}  
}
\caption{Azimuthally-averaged surface density as a function
  of radius at various times.   Radial distance is in units of $a$, surface density in units of
  $\Sigma_0$.  Time averages were performed over windows
  $\Delta t = 2000\mbh$ in size centered on different times, given in
  units of $10^4 \mbh$ in the legend.   (Left-to-right) \massratiolist. 
}
\label{fig:surfdens_over-time-mass-ratios}
\end{figure*}

As time progresses through each run, in all but \qone there is a strong convergence
in $\Sigma(r)$.   Much of this convergence takes place at times earlier than
the duration of \qone, indicating that time-steadiness is achieved more
rapidly with smaller $q$.   The steadiness of $\Sigma(r)$ at late times in
all runs is consistent with the near steady-state of enclosed mass shown in
Section~\ref{sec:enclosed-mass}.

During this convergence, the shape of $\Sigma(r)$ gradually changes,
and the nature of the converged shape is a function of $q$.  We show
$\Sigma(r) = \timesmooth{\Sigma(r,t)}{\Delta t = 2000\mbh}$ at a
number of times in each run in
Figure~\ref{fig:surfdens_over-time-mass-ratios}.  Several properties
show clear trends as the mass-ratio decreases.  First, the surface
density's peak broadens and its contrast with the surface density at
larger radius diminishes.  Second, the inner edge of the gap moves
inward in terms of $r/a$, consistent with the decline of the binary's
quadrupole moment, which destroys closed orbits.  This effect may also
be viewed as a weakening of the binary torques, which can repel
material outward. 
Previous Newtonian work (\cite{DOrazio13,Farris14b,Miranda2017}: 2-d
$\alpha$-viscosity hydrodynamics) and (\cite{Shi15}: MHD) found
a similar trend, while full GRMHD simulations around closer binaries
\citep{Gold14} showed little dependence of the circumbinary disk edge
position on mass ratio.

\begin{figure*}[htb]
\centerline{
\includegraphics[width=2\columnwidth]{\plotdir/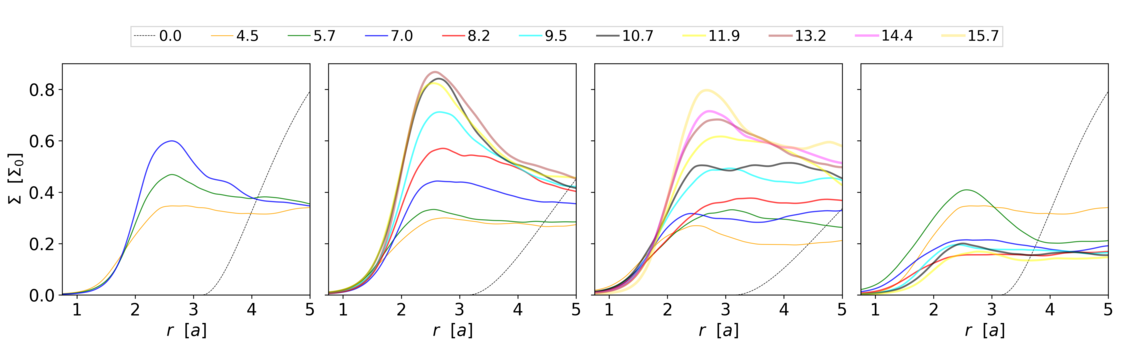}  
}
\caption{Temporal and azimuthal averaged surface density as a function
  of radius in units of $a$.  Time averages are performed over windows
  $\Delta t = 2000\mbh$ in size centered on different times, specified in
  units of $10^4 \mbh$ in each legend.  (Left-to-right) \magfluxlist. }
\label{fig:surfdens_over-time-states}
\end{figure*}

The \magfluxruns show a greater variety of behavior and, by this
measure, are slower to approach inflow equilibrium.  The larger
and more distant mass reservoirs of \medrun and \lrgrun result in
flatter $\Sigma(r)$ profiles at earlier times as mass more slowly
redistributes itself to smaller radii.  For instance, the local
maximum in $\Sigma(r)$ is apparent in \medrun and \lrgrun only after
$t \gtrsim 8\times10^4\mbh$, and appears to
converge to a steady value by $t \simeq  1.1\times10^5\mbh$ in both runs.
Also, the surface density's
local maximum in \medrun seems to be significantly narrower and larger
compared to its value at larger radii. This last distinction may be
the result of the inner portion of \lrgrun's accretion flow
equilibrating with its outer part; the difference between
$\Sigma(r=2.5a)$ and $\Sigma(r=5a)$ is much larger for \medrun, and
\lrgrun's enclosed mass at $r=3a,4a$ appears to be growing faster than
that of \medrun at late times.

The effect of the magnetic flux injection is apparent in $\Sigma(r,t)$
of \injectrun.   The perturbation creates a broad peak
in $\Sigma(r)$ just after the time of injection, $t=5.2\times10^4\mbh$.
As the perturbation enhances redistribution of gas and angular
momentum, the local maximum's relative amplitude decays
over time---as does the absolute magnitude of $\Sigma(r,t)$. 
Thus, as already noted, this run does not come particularly
close to inflow equilibrium.

\subsection{Torque Density}
\label{sec:torque-density-mass-ratios}

\begin{figure*}[htb]
\centerline{
  \begin{tabular}{c}
  \includegraphics[width=2\columnwidth]{\plotdir/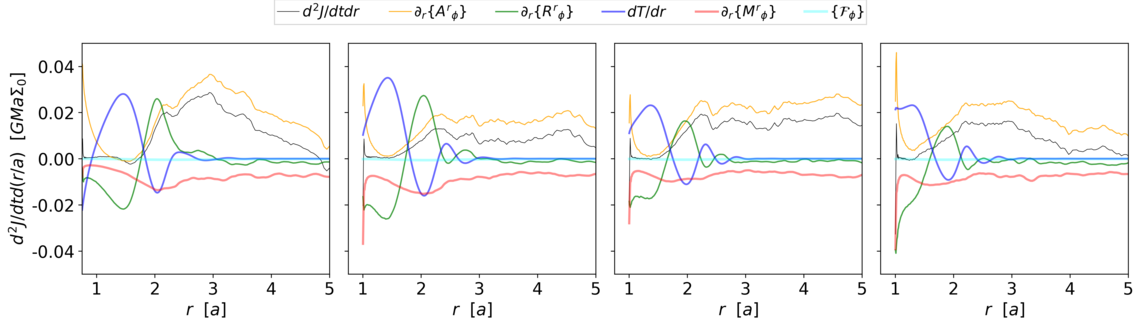}\\
  \includegraphics[width=2\columnwidth]{\plotdir/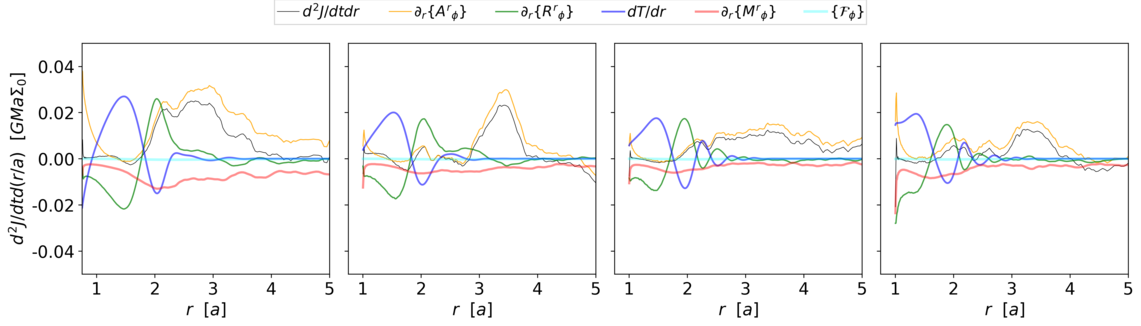}
  \end{tabular}
}
\caption{Contributions to the time-average radial distribution of
  $\partial_r \partial_t J$ (black) in the \massratioruns.  Shown are
  the radial derivatives
  of the Maxwell stress in the Eulerian frame ($\left\{ {M^r}_\phi
  \right\}$, red), the angular momentum flux due to shell-integrated
  Reynolds stress in the Eulerian frame ($\left\{ {R^r}_\phi
  \right\}$, green), and advected angular momentum ($\left\{
          {A^r}_\phi \right\}$ , gold).  Also shown are the torque
          densities per unit radius due to the actual binary spacetime
          ($dT/dr$, blue) and radiation losses
          ($\left\{\mathcal{F}_\phi\right\}$, cyan).  The net rate of
          change of angular momentum $\partial_r \partial_t J$ (solid
          black).  All quantities in the top (bottom) row plots
          are time-averaged over
          $40,000 < t/M< 76,000$ (last $30,000\mbh$ of evolution per run). 
          (Left-to-right) \massratiolist.
          Note that $\partial_r \left\{ {M^r}_\phi\right\}$,
          $\partial_r \left\{ {R^r}_\phi\right\}$ , $\partial_r
          \left\{ {A^r}_\phi \right\}$, and
          $\left\{\mathcal{F}_\phi\right\}$ have all been multiplied
          by a factor of $-1$ to match the sign they have in
          Eq.~\eqref{ang-mom-profile-4} so their curves add up to that
          of $\partial_r \partial_t J$. }
\label{fig:djdtdr-mass-ratios}
\end{figure*}

Several different mechanisms transport angular momentum within the
circumbinary disk.   Because angular momentum conservation
is broken by non-axisymmetry in the gravitational spacetime
spacetime, we discuss this issue here.

The total angular momentum $J$ is the
integral over the spatial volume of the time component of its
associated current, $j^\mu$: $J = \int j^t \, \sqrt{-g} \, dV$, where
$dV$ is the spatial volume component in the spacelike hypersurface
(e.g., $dr d\theta d\phi$).  We are interested in the azimuthal
component of the momentum, so the desired current is $j^\mu =
{T^\mu}_\nu \phi^\nu$, and $\phi^\nu = \left(\partial_\phi\right)^\nu
= \partial x^\nu/\partial \phi = [0,0,0,1]$ in spherical coordinates,
which is what we use.   

Radial transport of angular momentum can be traced through examination
of the several mechanisms contributing to the local rate rate of change of
angular momentum density, $d^2 J/dt dr$:
\beqa{
\partial_r \partial_t  J  & = & dT/dr  -  \left\{ \mathcal{F}_\phi \right\}
 -  \partial_r  \left\{ {T^r}_\phi \right\} \nonumber \\
& = & dT/dr  - \left\{ \mathcal{F}_\phi \right\}
 - \partial_r  \left\{   {M^r}_\phi \right\} \nonumber \\
&& - \partial_r  \left\{   {R^r}_\phi \right\}
 - \partial_r  \left\{   {A^r}_\phi \right\}
 \quad . 
\label{ang-mom-profile-4}
}
The radial density of gravitational torque is
\beq{
  \deriv{T}{r} = \frac{1}{2} \int T^{\mu \nu} \left( \partial_\phi g_{\mu \nu}\right)  \sqrt{-g} \, d\theta \, d\phi
\quad .   \label{torque-density}
}
The quantities ${M^r}_\phi$, ${R^r}_\phi$, and ${A^r}_\phi$
are---respectively---the Maxwell (MHD) stress, Reynolds stress, and
advected flux of angular momentum.   The Maxwell stress
${M^\mu}_\nu = 2p_m u^\mu u_\nu + p_m {\delta^\mu}_\nu - b^\mu b_\nu$ 
is the EM part of the stress-energy tensor, while the Reynolds stress and the
advected angular momentum flux sum to the hydrodynamic part:
$\left( {R^\mu}_\nu + {A^\mu}_\nu \right) = {{T_H}^\mu}_\nu = \rho h u^\mu u_\nu + p {\delta^\mu}_\nu$. 

The quantities $\left\{ {R^\mu}_{\nu} \right\}$ and $\left\{ {A^\mu}_{\nu} \right\}$ can be
separated by defining the advected flux in terms of the mean radial flow and then
subtracting it from the total hydrodynamic angular momentum flux:
\beq{ 
\left\{ {A^r}_\phi  \right\} \simeq \frac{\left\{ \rho \ell \right\}  \left\{ \rho h u^r \right\} }{\left\{ \rho \right\}}
\,. \label{advected-flux}
}\beq{
\left\{ {R^r}_\phi \right\} 
\ = \ \left\{ \rho h \, \delta u^r \, \delta u_\phi  \right\}
\ \simeq \  \left\{ {{T_H}^r}_\phi \right\} - \left\{ {A^r}_\phi \right\} 
\label{reynolds-derivation-11}
}
Here $\ell \equiv - u_\phi / u_t$.  

We show each of these contributions separately in
Figure~\ref{fig:djdtdr-mass-ratios}.  The contributions are displayed averaged over
two different epochs: an earlier period ($4\times10^4 < t/M<
7.6\times10^4$) to compare with the final part of \qone, and the last
$3\times10^4\mbh$ of each run in order to illustrate how the system evolves
as it nears inflow equilibrium.  

Several qualitative conclusions can be drawn from these figures.
First, all the runs of the \massratioruns reach a steady-state with respect
to angular momentum transport at $r \gtrsim 3a$ by the time of the
later period, evidenced by the total angular momentum gradient
lying close to zero.   The magnitudes of all the contributions diminish slightly in time,
with the largest decrease in the advected and magnetic contributions.

Second, there is only weak dependence on mass-ratio.
The peak of the gravitational torque density,
$dT/dr$, moves inward as $q$ decreases.   With only
a few exceptions where both are small, the radial gradient of the Reynolds stress
contributes to the total torque so as to cancel the gravitational torque; in other words,
when gravitational torque adds angular momentum to the fluid in a grid-cell, Reynolds
stress carries it away.

This figure also illustrates the transition
from linear gravitational torques to nonlinear.  At late times (bottom
row of this figure), as $q$ rises from 0.1 to 1,
the damped sinusoidal oscillations in both gravitational torque density and
Reynolds stress as functions of radius flatten out into low-amplitude plateaus
at the third extremum.  This plateau feature resides at the location of the
lump and demonstrates that the response of the circumbinary disk matter
to external torques can no longer be described by linear perturbation theory
when $q \gtrsim 1/4$.

On the other hand, the Maxwell stress is consistently close to 
uniform spatially for all $q$; the magnitude of this spatially-uniform
stress is almost the same for all $q \leq 0.5$.  The degree to
which the Maxwell stress maintains the same constant value over all
mass ratios suggests that its magnitude reflects the asymptotic behavior of the MRI.  As shown in \cite{Noble12} for \qone, and found
in the other runs but not shown here, the plasma
$\beta = p / p_m$ exhibits a local maximum at $r \simeq 2.5a$; this maximum is more pronounced for larger $q$.
The two trends together imply that larger $q$ leads to higher pressure at the radius of the surface density maximum.  This greater pressure may be the end-result of the stronger
gravitational torques associated with higher $q$ doing more mechanical work, and the eventual dissipation of this work into heat, rather than a
loss of magnetic field intensity.

\begin{figure*}[htb]
\centerline{
  \begin{tabular}{c}
  \includegraphics[width=2\columnwidth]{\plotdir/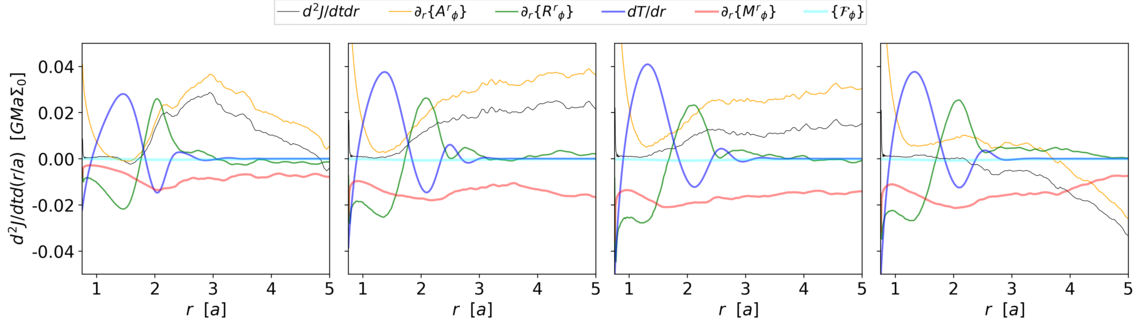}\\
  \includegraphics[width=2\columnwidth]{\plotdir/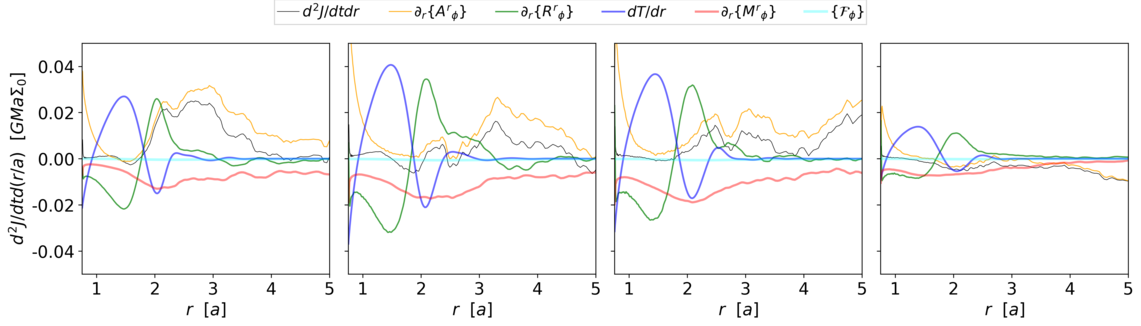}
  \end{tabular}
}
\caption{Contributions to the time-average radial distribution of
  $\partial_r \partial_t J$ (black) in the \magfluxruns.
  Shown are the radial derivatives
  of the Maxwell stress in the Eulerian frame ($\left\{ {M^r}_\phi
  \right\}$, red), the angular momentum flux due to shell-integrated
  Reynolds stress in the Eulerian frame ($\left\{ {R^r}_\phi
  \right\}$, green), and advected angular momentum ($\left\{
          {A^r}_\phi \right\}$ , gold).  Also shown are the torque
          densities per unit radius due to the actual binary potential
          ($dT/dr$, blue) and radiation losses
          ($\left\{\mathcal{F}_\phi\right\}$, cyan).  The net rate of
          change of angular momentum $\partial_r \partial_t J$ (solid
          black).  All quantities in the top (bottom) row plots
          are time-averaged over
          $40,000 < t/M< 76,000$ (last $30,000\mbh$ of evolution per run). 
          (Left-to-right) \magfluxlist.
          Note that $\partial_r \left\{ {M^r}_\phi\right\}$,
          $\partial_r \left\{ {R^r}_\phi\right\}$ , $\partial_r
          \left\{ {A^r}_\phi \right\}$, and
          $\left\{\mathcal{F}_\phi\right\}$ have all been multiplied
          by a factor of $-1$ to match the sign they have in
          Eq.~\eqref{ang-mom-profile-4} so their curves add up to that
          of $\partial_r \partial_t J$.
          }
\label{fig:djdtdr-states}
\end{figure*}

The angular momentum budget for the \magfluxruns is shown in
Figure~\ref{fig:djdtdr-states}.   Like the \massratioruns, the fact
that $d^2J/dt dr$ settles toward the zero line at late times  gives
strong evidence of approach to a steady-state.   There is also
a resemblance to the \massratioruns in the sense that, like \qone and \qtwo,
the \magfluxruns show an absence of the third peak in the Reynolds stress at late
times. 

However, there
is more contrast between these runs than those in the \massratioruns .
  During the earlier averaging period,
$d^2J/dt dr$ is far from zero throughout
the circumbinary disk in \origrun, \medrun, and \lrgrun, indicating
that this is a transient phase in mass/magnetic-flux
redistribution for all three. In \injectrun, $d^2J/dt dr \simeq 0$ for
$r \lesssim 3.5a$, but grows rapidly at larger radius, showing that
this run reached a steady-state in its inner regions more rapidly than
the others, but in this time-span is evolving rapidly at larger radii.
At later times, all three new runs come much closer to equilibrium in
their angular momentum evolution.

To close this section on axisymmetric properties, we remark on
how the non-axisymmetric lump can influence azimuthally-averaged properties
such as the vertically-integrated magnetic stress and the MRI quality factors.
The largest value of the former over  the entire radial extent of the circumbinary
disk is found at the radial location of the lump, even though the minima for the
latter are found at the
$\left(r,\phi\right)$ locations of the lump (see
Appendix~\ref{app:mri-resolution}).
To explain this diminution in MRI quality, we point out that the
magnetic stresses of \origrun, \medrun, and \lrgrun all agree at
$r=5a$, suggesting that the variations between those runs neither
strengthen nor weaken the field in the outer disk.  Nonetheless,
in \medrun and \lrgrun, the stresses at $r \simeq 2a$, i.e., the lump
region, are even larger than in \origrun.   This fact suggests that most of the
degradation in MRI quality in these runs must be due to increased density
in the lump region.

\begin{figure*}[htb]
\centerline{
  \includegraphics[width=2\columnwidth]{\plotdir/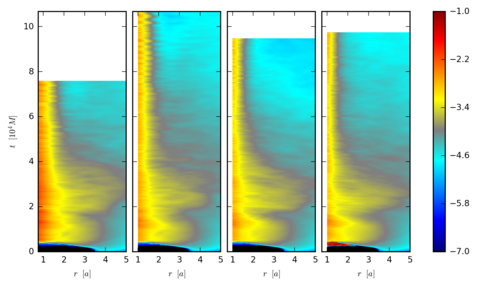}
}
\caption{Log$_{10}$ of the ratio of shell-averaged
  Maxwell stress, ${M^r}_\phi$, to shell-averaged mass density, $\shellavg{\rho}$, versus radius and
  time, shown in units of binary separation and total BH mass,
  respectively. The scale is shown in the color bar. The ranges of
  time and radius used in the plots cover the full extents of each
  simulation. (Left-to-right) \massratiolist.
\label{fig:maxwell-to-density-spacetime-mass-ratio}}
\end{figure*}

\begin{figure*}[htb]
\centerline{
  \includegraphics[width=2\columnwidth]{\plotdir/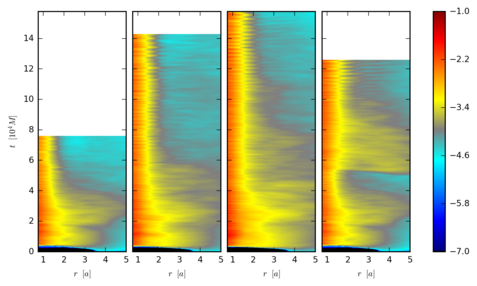}
}
\caption{Log$_{10}$ of the ratio of shell-averaged
  Maxwell stress, ${M^r}_\phi$, to shell-averaged mass density, $\shellavg{\rho}$, versus radius and
  time, shown in units of binary separation and total BH mass,
  respectively. The scale is shown in the color bar. The ranges of
  time and radius used in the plots cover the full extents of each
  simulation. (Left-to-right) \magfluxlist.
\label{fig:maxwell-to-density-spacetime-states}}
\end{figure*}

In order to explore how magnetic stress may influence
lump dynamics and evolution, it is useful to define a measure of the magnitude of
the magnetic stress per unit mass, which we will call ${W^r}_\phi$
following \cite{BH98}:
\beq{
  {W^r}_\phi =
  \frac{\shellint{{M^r}_\phi}}{\shellint{\rho}} \quad
  . \label{maxwell-stress-per-unit-mass}
}
This quantity for the \massratioruns and the \magfluxruns is shown in
Figure~\ref{fig:maxwell-to-density-spacetime-mass-ratio} and
Figure~\ref{fig:maxwell-to-density-spacetime-states}, respectively.
In every run of the mass-ratio series, ${W^r}_\phi$ at radii $r \gtrsim 2a$
drops abruptly
by about a factor of 4 at a time $\approx 40,000M$.  Particularly
for low $q$, this drop begins at large radius and only then extends
inward.  The evolution of ${W^r}_\phi$ in the magnetic flux series
is very different because we deliberately manipulated the magnetic
flux available.

For those runs with a lump, we find that once the specific
  magnetic stress drops to ${W^r}_\phi \lesssim 10^{-4}$ the lump
  appears when one uses the criteria described in
  Section~\ref{sec:non-axisymm-struct}. The significance of this value
  will be discussed in Section~\ref{sec:orig-lump-accr}.

\section{Non-axisymmetric Structure}
\label{sec:non-axisymm-struct}

\subsection{Lump Amplitude}
\label{sec:lump-amplitude}

\begin{figure}[htb]
\centerline{
  \includegraphics[width=\columnwidth]{\plotdir/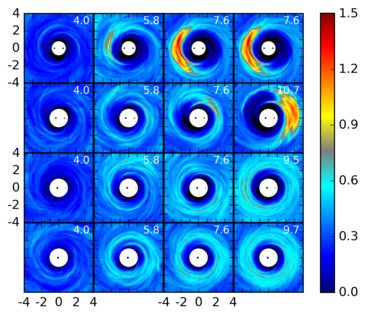}
}
\caption{Color contours of surface density in units of $\Sigma_0$ as a
  function of radius and azimuthal angle, ,
  i.e. $\Sigma(r,\phi)/\Sigma_0$, at four different times indicated by
  the number in the upper-right corner of each frame in units of $10^4
  \mbh$.  The first three times were chosen to  span
  \qone's secularly-evolving state; the time of the right-most column
  is the last time in each run.  (Top to bottom)
  \massratiolist.  (Right) linear color scale used in
  all frames.
\label{fig:surfdens_equatorial-mass-ratios}}
\end{figure}

\begin{figure}[htb]
\centerline{
  \includegraphics[width=\columnwidth]{\plotdir/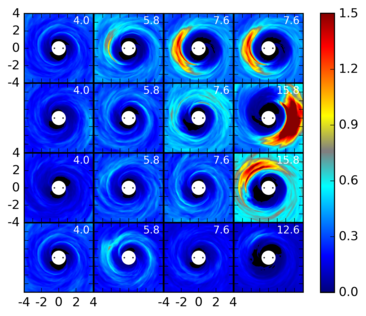}
}
\caption{Color contours of surface density in units of $\Sigma_0$ as a
  function of radius and azimuthal angle, ,
  i.e. $\Sigma(r,\phi)/\Sigma_0$, at four different times indicated by
  the number in the upper-right corner of each frame in units of $10^4
  \mbh$.  The first three times were chosen to  span
  \qone's secularly-evolving state; the time of the right-most column
  is the last time in each run.  (Top to bottom)
  \magfluxlist. (Right) linear color scale used in
  all frames.
\label{fig:surfdens_equatorial-states}}
\end{figure}

Although
Figures~\ref{fig:surfdens_over-time-mass-ratios}~and~\ref{fig:surfdens_over-time-states}
illustrate well the azimuthally-averaged pileup of material at the edge of the gap for all
runs, they lack information about non-axisymmetric structure.
In particular, they say nothing about the lump feature \cite{MM08,Shi12,Noble12},
which can affect the electromagnetic signal both by the dissipation associated with
it directly and by its modulation of the accretion rate. In order to investigate the 
non-axisymmetric structure of the flow, in
Figures~\ref{fig:surfdens_equatorial-mass-ratios}~and~\ref{fig:surfdens_equatorial-states}
we  plot the distribution of the surface density, $\Sigma(r,\phi)$ at evenly spaced
intervals over the secularly-evolving period of the runs.

In all cases with a lump, we find that the overdensity region spans
$\pi/3 \lesssim \delta\phi_\mathrm{lump} \lesssim \pi$ in azimuthal angle,
and a radial extent of $0.1 \lesssim \delta r_\mathrm{lump} \lesssim
a$.   In many cases, the density contrast between the lump and its
surroundings is quite large.

In the \massratioruns, the development of an azimuthally asymmetric
overdensity is obvious only in runs \qone and \qtwo, with the latter run showing weaker
development at all the times shown in this figure.
The lump does not appear at any time in either \qfive or \qten.

On the other hand, all four runs in the \magfluxruns show lumps in at least one of
the snapshots shown.
The lumps develop more slowly in tori extending to larger radius such as \medrun
and \lrgrun.  When their lumps form, however, the peak surface density in each is
significantly greater than in \origrun.  The impact of additional mass-supply is evident.  

The images of \injectrun look different than the others, but this is a visual
artifact of the burst of accretion triggered by the injected magnetic
field. As a result, substantially less mass remains in the disk.  Nonetheless,
the {\it contrast} between the surface density of the lump and the azimuthally-averaged
surface density at late times is comparable to that in the other runs
(see Figure~\ref{fig:lump-relative-mode-amplitudes-spacetime-init-states}
for a clearer view of this contrast).
Note that at $t=5.8\times10^4 \mbh$ one may still see the remnant of the $m=1$ lump
structure created in \origrun before the extra magnetic flux was injected in \injectrun.

\begin{figure*}[htb]
\centerline{
  \includegraphics[width=2\columnwidth]{\plotdir/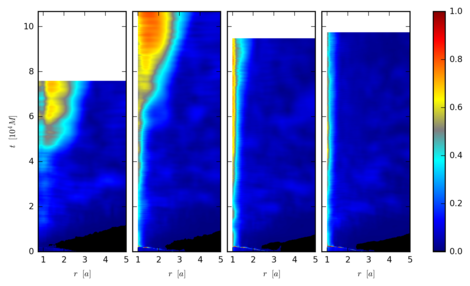}
}
\caption{The ratio of the $m=1$ mode to the $m=0$,
$\timesmooth{A_1}{r,t,\Delta t}/\timesmooth{A_0}{r,t,\Delta t}$, as a function of radius
and time for the \massratioruns. Here the smoothing period slides along
$t$ and has duration $\Delta t = 2\tlump$. (Left-to-right) \massratiolist .
 }
\label{fig:lump-relative-mode-amplitudes-spacetime-mass-ratios}
\end{figure*}

\begin{figure*}[htb]
\centerline{
  \includegraphics[width=2\columnwidth]{\plotdir/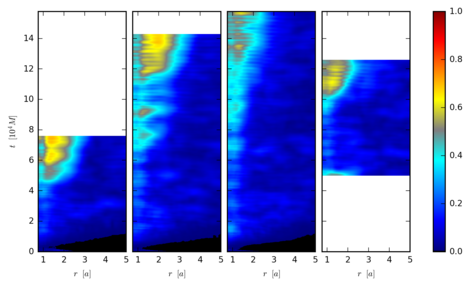}
}
\caption{The ratio of the $m=1$ mode to the $m=0$,
$\timesmooth{A_1}{r,t,\Delta t}/\timesmooth{A_0}{r,t,\Delta t}$, as a function of radius
and time for the \magfluxruns. Here the smoothing period slides
along $t$ and has duration $\Delta t = 2\tlump$.   (Left-to-right) \magfluxlist.}
\label{fig:lump-relative-mode-amplitudes-spacetime-init-states}
\end{figure*}

To quantify the surface density contrast between the lump
and its surroundings, we compute the Fourier transform ${\mathcal B}_m$ of
$\theta$-integrated $\rho \sqrt{-g}$ with respect to $\phi$ in the coordinate
frame. We call its absolute magnitude $A_m$, with the definitions
\beq{ A_m(r,t) =
  \left| \mathcal{B}_m(r,t) \right| \quad , \quad \mathcal{B}_m(r,t) = \shellint{ \rho  e^{i m \phi}}
  \quad .
  \label{mode-amplitudes}
}
The phase of the Fourier modes indicates the azimuthal location
of the lump:
\beq{ \phase_m(r,t) = 
  \mathtt{ArcTan}\left(-\operatorname{Im}(\mathcal{B}_m(r,t)), \operatorname{Re}(\mathcal{B}_m(r,t))\right) \quad
  \label{mode-phase}
}
where our $\mathtt{ArcTan}(y,x)$ function returns the angle between the $y=0, x>0$ line and the line connecting the point $(x,y)$ and the origin.

As shown in Figure~\ref{fig:lump-relative-mode-amplitudes-spacetime-mass-ratios},
 $A_1(r,t)/A_0(r,t)$ for \qone and \qtwo increases substantially over time.
The $m=1$ mode amplitude is strongest in the
  accretion stream region, but grows significantly in the region of the lump proper, $r \approx 2.5a$ as well.   That both the inner region of the circumbinary disk and the stream
  region develop the same sort of asymmetry is no coincidence.   If there were no disk asymmetry, the stream region would be modulated strongly for $m=2$, not $m=1$; that the streams also have $m=1$ character is a sign that the accretion
  streams originate in the lump.  The amplitude of the $m=1$ mode in the gap is larger than in the disk because the streams grow narrower and denser as they
  fall toward the nearest black hole and the remainder of the gap has such low density.
By contrast, \qfive and \qten show almost no signs of growth in the relative amplitude of the  $m=1$ mode.

All the \magfluxruns runs show enhancements of $A_1/A_0$ similar
to those seen in \origrun and \qtwo,  but at
rather later times (the brief appearance of
significant $A_1/A_0$ at the beginning of \injectrun is the remnant
of the lump in \origrun as it is destroyed by the injection of
magnetic flux).

\begin{figure*}[htb]
\centerline{
  \includegraphics[width=2\columnwidth]{\plotdir/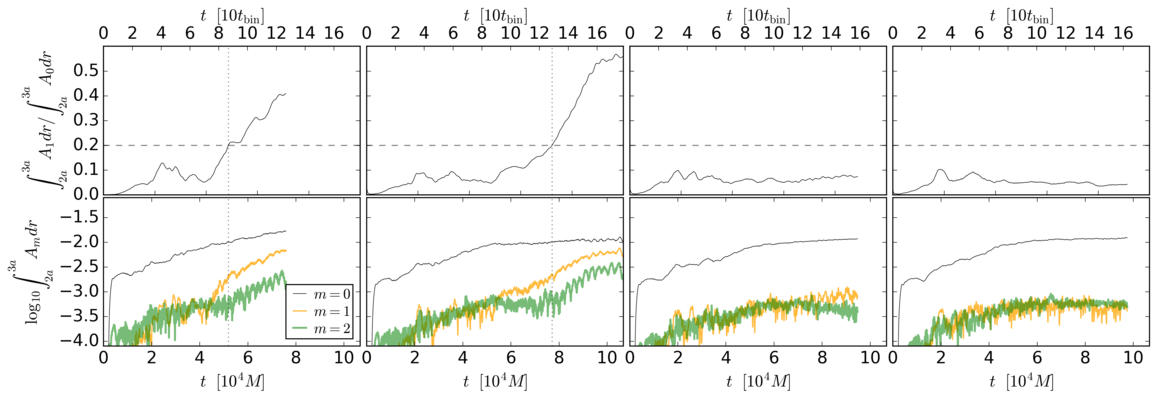}
}
\caption{Lump criterion, $\critlump(t)$, (top rows) used to determine
  $\Tlump$, and the $m=0,1,2$ mode amplitudes integrated over the lump
  region (bottom rows) for the mass-ratio simulations.  The horizontal
  dashed line indicates the threshold $\critlump=0.2$ above which we
  recognize the presence of an overdensity; the vertical lines denote
  the time, $\Tlump$, at which it first satisfies this criterion.
  (Left-to-right) \massratiolist. }
\label{fig:lump-criteria-mass-ratios}
\end{figure*}
 
\begin{figure*}[htb]
\centerline{
  \includegraphics[width=2\columnwidth]{\plotdir/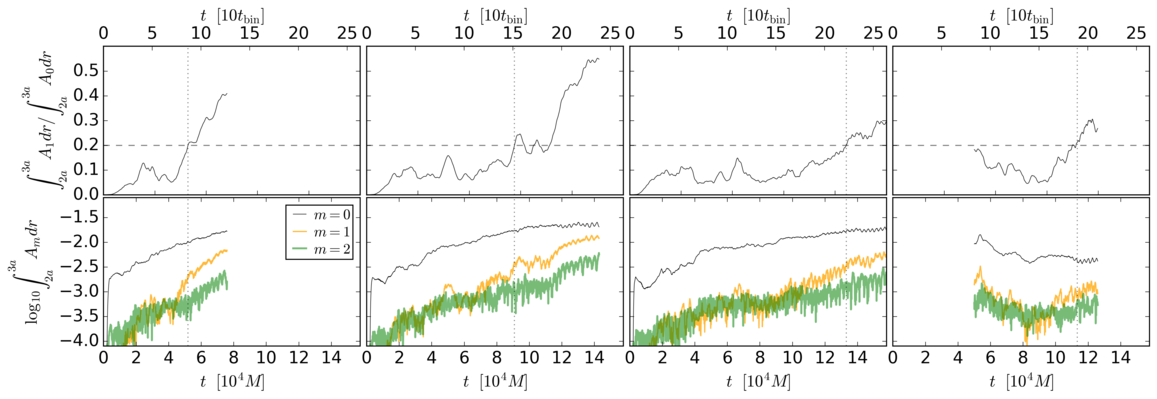}
}
\caption{Lump criterion, $\critlump(t)$, (top rows) used to determine
  $\Tlump$, and the $m=0,1,2$ mode amplitudes integrated over the lump
  region (bottom rows).  The horizontal
  dashed line indicates the threshold $\critlump=0.2$ above which we
  recognize the presence of an overdensity; the vertical lines denote
  the time, $\Tlump$, at which it first satisfies this criterion.
  (Left-to-right) \magfluxlist.}
\label{fig:lump-criteria-init-states}
\end{figure*}

Interpreting a large relative mode strength of the $m=1$ mode
as a signature of the lump, we determine the onset of the lump using the criterion
\beqa{
  \critlump(t) =
  \frac{ \int_{2a}^{3a} \timesmooth{A_1(r,t)}{2\tlump} \, dr}{\int_{2a}^{3a} \timesmooth{A_0(r,t)}{2\tlump} \, dr}
  > 0.2.
  \label{criterion} 
}
We plot $\critlump(t)$ for each run  in both series, along with the $m=0,1,2$ mode
amplitudes integrated over $r \in \left[ 2 a , 3 a \right]$, in
Figures~\ref{fig:lump-criteria-mass-ratios}-\ref{fig:lump-criteria-init-states}.
Satisfaction of the criterion coincides with the time when the amplitude of the $m=1$
mode begins to rise above that of the $m=2$ mode and approaches the $m=0$
mode strength.  We define $\Tlump$ to be the time the lump criterion is first satisfied.
The measured values of $\Tlump$ for all lump-forming runs are presented in Table~\ref{tab:run-characteristics}.

Similar behavior is seen in the \magfluxruns.  We find that each run
in this series satisfies the criterion within its duration. Consistent with the surface
density plots, the lump forms later for \medrun and
\lrgrun. The $m=1$ amplitude of \medrun reaches more than half that of
its $m=0$ mode, while in \lrgrun this ratio crosses
the threshold of 0.2 near the end of the simulation and reaches 0.3 at
the very end. In \injectrun, the run begins with the
decay of the existing lump from \origrun, but the lump recovers and
crosses the threshold later.

An important characteristic of the lump is its phase
  coherence. In order to quantify the instantaneous phase of the lump,
  $\phaselump(t)$ and its associated orbital frequency,
  $\omegalump(t)$, we rely on our means of calculating the amplitude
  of the overdensity Eq.\eqref{mode-amplitudes}.  As we want to track the
  lump, which resides close to the cavity's edge, we first integrate
  the density's $m=1$ Fourier amplitude over the radial extent of the lump:
  \beq{
    \tilde{\mathcal{B}_1}(t) \equiv \int_{2a}^{4a} \mathcal{B}_1(r,\phi,t) \, dr  \label{mode-amplitude-1}
  }
  which is then immediately used to find $\phaselump(t)$:
  \beq{
    \phaselump(t) \equiv \mathtt{ArcTan}\left(-\operatorname{Im}(\tilde{\mathcal{B}}_1(t)), \operatorname{Re}(\tilde{\mathcal{B}}_1(t))\right) \quad  \quad . \label{phase-lump} 
  }
 The instantaneous orbital frequency of the lump,
  $\omegalump(t)$, is simply the time derivative of the phase:
\beq{
   \omegalump(t) = \deriv{\phaselump(t)}{t} \quad . \label{omega-lump} 
}
When referenced without an argument, $\omegalump$ is to
be interpreted as the time average of $\omegalump(t)$ over the period
$\Tlump < t < \tend$ for specified run.
Table~\ref{tab:run-characteristics} shows the values of $\omegalump$
for each run.

\subsection{Eccentricity}
\label{sec:eccentricity}

\begin{figure*}[htb]
\centerline{
\begin{tabular}{cc}
  \includegraphics[width=\columnwidth]{\plotdir/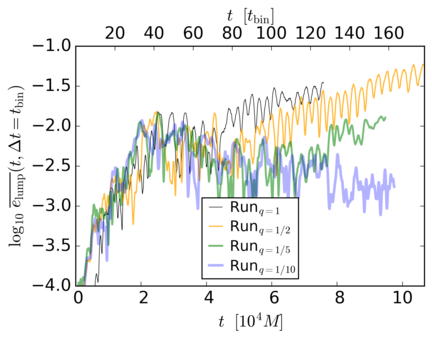}
& \includegraphics[width=\columnwidth]{\plotdir/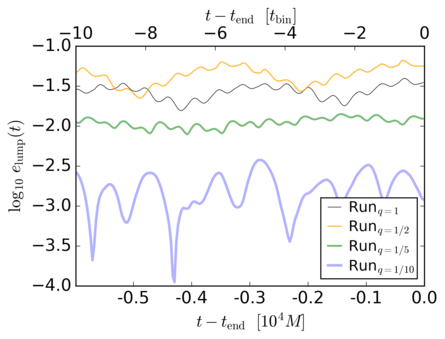}
\end{tabular}
}
\caption{The lump's eccentricity smoothed over $\Delta t = \tbin$, 
  $\timesmooth{\elump}{t,\tbin}$ (left); $\elump(t)$
  over the last 10 orbits of each simulation (right); $\tend$
  represents the final time of each simulation of the \massratioruns.}
\label{fig:lump-eccentricity-mass-ratios}
\end{figure*}

\begin{figure*}[htb]
\centerline{
\begin{tabular}{cc}
  \includegraphics[width=\columnwidth]{\plotdir/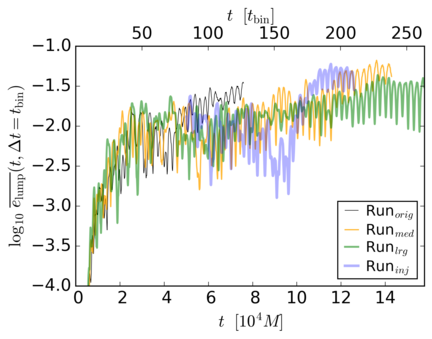}
& \includegraphics[width=\columnwidth]{\plotdir/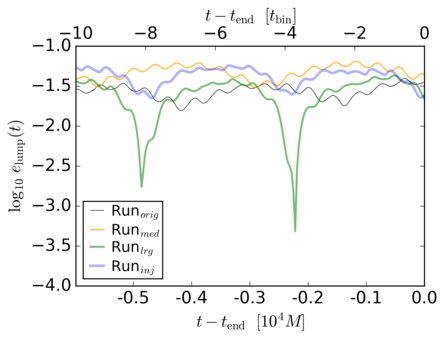}
\end{tabular}
}
\caption{The lump's eccentricity smoothed over $\Delta t = \tbin$, 
  $\timesmooth{\elump}{t,\tbin}$ (left); $\elump(t)$
  over the last 10 orbits of each simulation (right); $\tend$
  represents the final time of each simulation of the \magfluxruns.}
\label{fig:lump-eccentricity-init-states}
\end{figure*}

Even if the circumbinary gas begins the simulation
on circular orbits, previous investigations reported that it acquires
non-trivial levels of eccentricity over time
\cite{MM08,Shi12,Farris14a,DOrazio2016,Miranda2017}.  At late times, the azimuthally-averaged
eccentricity rises sharply just inside the cavity wall and decays exponentially
outward $\approx \exp{(-r/a)}$ \cite{MM08,Shi12}.
Previous MHD simulations reported smaller eccentricities than
viscous hydrodynamics simulations, but it is unclear if the MHD simulations
reached a true steady state in terms of eccentricity.  We explore here how our
simulations compare with previous work and how eccentricity is
associated with the growth and strength of the lump.

We define the eccentricity of a disk annulus at radius $r$ in a way
analogous to the Newtonian expressions used in \cite{MM08,Shi12}, but
expressed in terms of the $4$-velocity and the metric:
\beq{
  e(r,t) =  \frac{\left|\shellint{\rho u^r e^{i \phi}}\right| }{
    \shellint{ \rho  r  u^\phi}  } .
  \label{eccentricity-annulus}
}
We further define the quantity $\elump$ to be the eccentricity of the
region $2 a < r < 4 a$:
\beq{
  \elump(t) = \frac{\left| \int^{4a}_{2a} dr \shellint{\rho u^r e^{i \phi} } \right| }{
    \int^{4a}_{2a} dr \shellint{\rho r u^\phi } } \quad . 
  \label{elump}
}
In order to accentuate variability or trends occurring at longer
time scales, we display $\elump(t)$ in
Figures~\ref{fig:lump-eccentricity-mass-ratios},\ref{fig:lump-eccentricity-init-states}
smoothed over a binary orbit: $\timesmooth{\elump}{t,\tbin}$,
calculated using Eq.~\eqref{tsmooth-def}.

As shown in Figures~\ref{fig:lump-eccentricity-mass-ratios} and
\ref{fig:lump-eccentricity-init-states}, in all cases the inner disk
eccentricity grows exponentially during the early development of
the circumbinary disk.  However, once it reaches $\sim 10^{-2}$, further
growth is a function of mass-ratio and magnetic flux. 

For fixed magnetic properties, but varying mass-ratio, the eccentricity
decreases slightly during the first $\sim 10^4\mbh$ after rising to
$\sim 10^{-2}$.  When the mass-ratio takes its maximum value, i.e.,
$q=1$, the eccentricity then renews its exponential growth, but at
a slower rate.  For smaller values of $q$, the growth is delayed
longer, and therefore begins from a lower level.  However, once
begun, growth proceeds at roughly the same rate for all values of $q$.
There is, however, one possible exception: for $q=0.1$, the slow
decline in eccentricity  runs all the way to the end of the simulation.
We cannot say whether the eccentricity might begin growing at still
later times.

On the other hand, for fixed mass-ratio and varying
magnetic properties, the development of eccentricity is very similar
in all the runs of this series.  The smoothed form of the eccentricity
history, $\timesmooth{\elump}{t,\tbin}$, ultimately reaches the same
value in each run to within $\sim 20\%$.
The only significant contrast between them occurs in \injectrun,
where the eccentricity drops sharply when the additional  magnetic flux
is added and the lump temporarily dissolves.  When the lump returns,
this run, too, returns to the common path.

The range of eccentricities we find all lie within the range of eccentricities observed
in another GRMHD series \citep{LopezArmengol2021} and about $50\%$
smaller than that reported in a Newtonian MHD study \citep{Shi12}.
Our results are also in agreement with 2-d VH results 
\citep{DOrazio2016},
and some 3-d SPH studies show a similar trend with $q$ \citep{Ragusa2020}. 

Generally speaking, in all the runs that exhibit a lump as
determined by our lump criterion, $\timesmooth{\elump}{t,\tbin}$,
grows as the $m=1$ density mode amplitude, $A_1$, grows. 
In both series, in each run exhibiting a lump, 
$\timesmooth{\elump}{t,\tbin}$ begins its second period of exponential growth approximately
$10^4\mbh$ before satisfying the lump criterion.
Moreover, the radial profiles are also quite similar to one another:
the time-averaged radial eccentricity profile for all cases declines
exponentially with increasing radius from $r \approx 1$ to $r \approx 3$ with
an $e$-folding scale consistently $\approx a$.

The time scales of variability seen in $\elump(t)$ for each run of the
\magfluxruns are also similar.  Each run exhibits a low-frequency
oscillation at $\omegalump$, the orbital frequency of the lump, and a
carrier signal at twice the beat frequency,
$2\left(\Omegabin-\omegalump\right)$.   The \massratioruns show more differences in their
  $\elump(t)$ variability than do the \magfluxruns . Whereas \qone and \qtwo exhibit variability
  at these two frequencies,  $\elump(t)$ in \qfive fluctuates at
  $2\Omegabin$ and $\elump(t)$ in \qten only at $\Omegabin$. 
  The weakness of the lump
  in \qfive and \qten explains their not varying at the beat
  frequency; the lower frequency variability of \qten is due to
  the fact that the secondary BH dominates the gravitational torque in the
  would-be lump region at this small mass ratio.

\subsection{Accretion Streams and Variability}
\label{sec:accr-stre-accr}

We ultimately aim to provide a quantitative model of how electromagnetic emission
depends on $q$ so that system parameters may be derived from observables. 
The bolometric luminosity $L$ is the simplest of observable measures.
Just as in \cite{Noble12}, we calculate $L$ by integrating the local 
cooling rate, $\mathcal{L}$, in the Eulerian frame over the numerical domain:
\beq{
  L(t) = \int \mathcal{L} \, u_t \, \sqrt{-g} \, dr \, d\theta \, d\phi
\quad .   \label{lightcurve-integral}
}
In such a method, Doppler and gravitational shifts are ignored, but the magnitude
of their effect is smaller than other uncertainties.
Each simulation is cooled toward the same target entropy, at the same cooling 
time scale, starting from tori at the same initial scale height ($H/R = 0.1$ at the 
pressure maximum) and target entropy.  
Hence, any changes in $L$ should be the result of the mass 
ratio or magnetic flux distribution, modulo statistical fluctuations.


\begin{figure*}[htb]
\centerline{
\includegraphics[width=2\columnwidth]{\plotdir/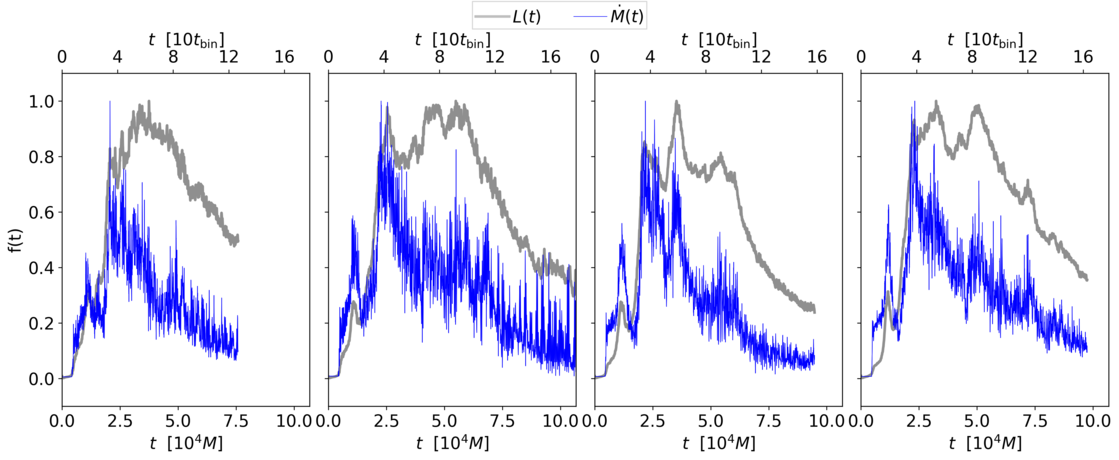}
}
\caption{Light curves and accretion rates, $\dot{M}(r=a,t)$, over each
  simulation's full extent. Each curve is normalized by its peak
  value.
  (Left-to-right) \massratiolist.
\label{fig:light-mdot-curves-mass-ratios}}
\end{figure*}

\begin{figure*}[htb]
\centerline{
\includegraphics[width=2\columnwidth]{\plotdir/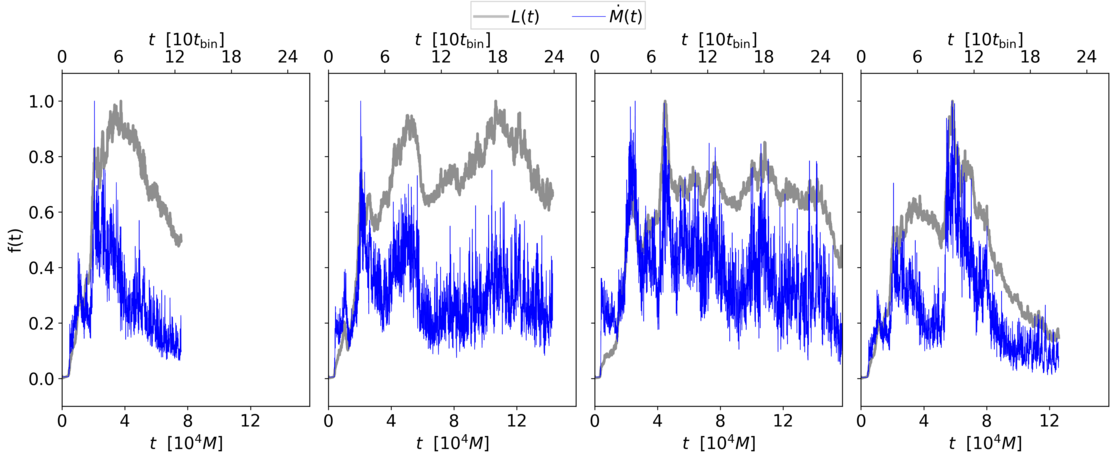}
}
\caption{Light curves and accretion rates, $\dot{M}(r=a,t)$, over each simulation's full extent. Each curve is normalized
  by its peak value. 
(Left-to-right)   \magfluxlist.
\label{fig:light-mdot-curves-init-states}}
\end{figure*}

\begin{figure*}[htb]
\centerline{
\includegraphics[width=2\columnwidth]{\plotdir/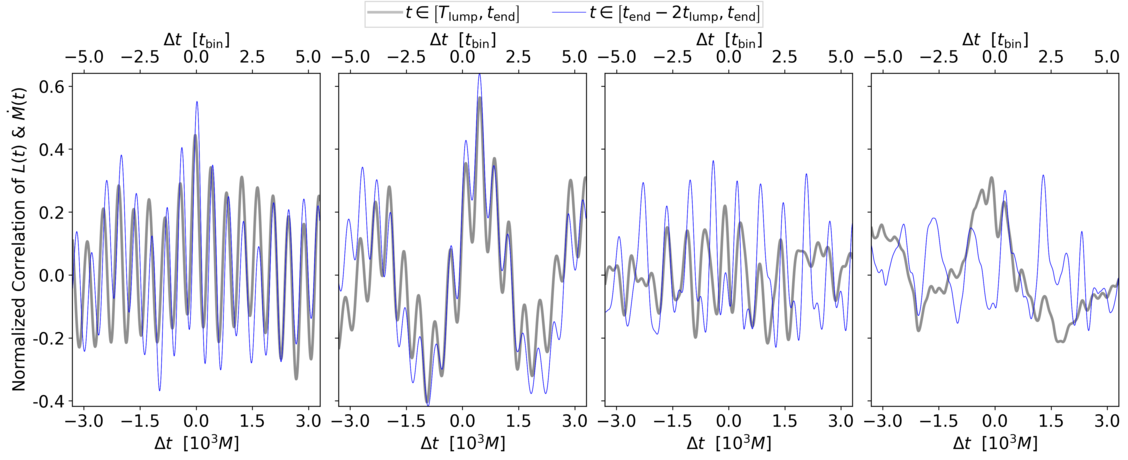}
}
\caption{Normalized correlations between the light curves, $L(t)$ and accretion rates, $\dot{M}(r=a,t)$, calculated
  since the onset of the lump (grey curves) or the last 2 periods of the lump's orbit in a simulation (blue curves).
  The correlations are plotted versus the lags, and are calculated using Eq.~\eqref{correlation-eq}. 
  A $5^\mathrm{th}$-order polynomial fit to each curve has been removed
  prior to calculating the correlation.  Each plot is displayed over a span of lag time approximately equal to $3\tlump$.
  (Left-to-right)  \massratiolist.
\label{fig:light-mdot-curves-corr-mass-ratios}}
\end{figure*}

\begin{figure*}[htb]
\centerline{
\includegraphics[width=2\columnwidth]{\plotdir/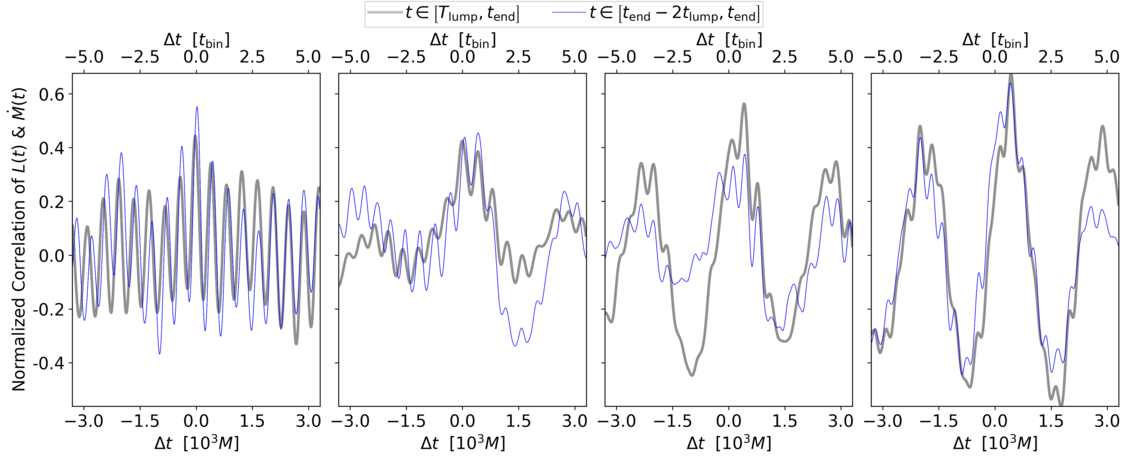}
}
\caption{Normalized correlations between the light curves, $L(t)$ and accretion rates, $\dot{M}(r=a,t)$, calculated
  since the onset of the lump (grey curves) or the last 2 periods of the lump's orbit in a simulation (blue curves).
  The correlations are plotted versus the lags, and are calculated using Eq.~\eqref{correlation-eq}. 
  A $5^\mathrm{th}$-order polynomial fit to each curve has been removed
  prior to calculating the correlation. Each plot is displayed over a span of lag time approximately equal to $3\tlump$.
  (Left-to-right)  \magfluxlist.
\label{fig:light-mdot-curves-corr-init-states}}
\end{figure*}


\begin{figure*}[htb]
\centerline{
\includegraphics[width=2\columnwidth]{\plotdir/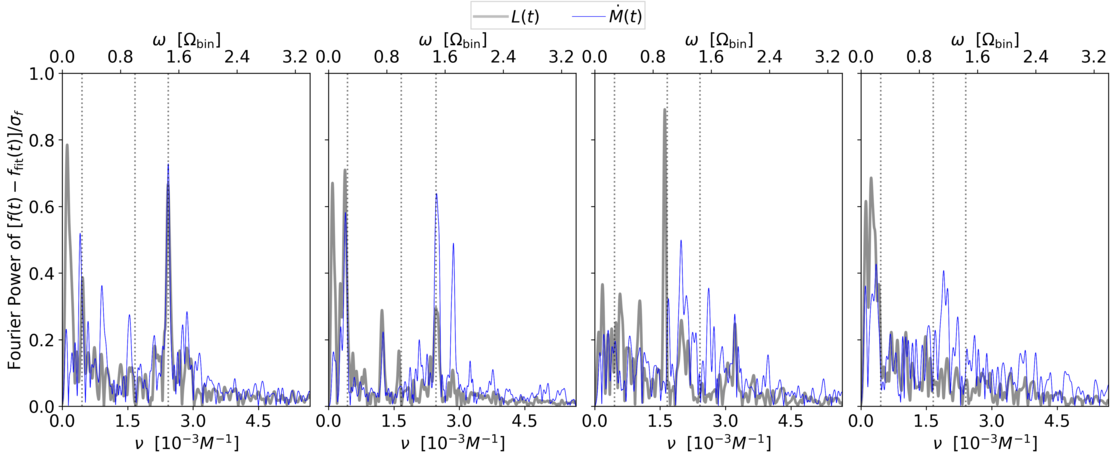}
}
\caption{Fourier power spectra of the light curves and accretion
  rates, $\dot{M}(r=a,t)$, including only times $t > \Tlump$.  For
  those runs with no observed lump, we use the simulation's last
  $2.5\times 10^4 \mbh$ of time.  Before performing the Fourier power
  spectrum, the function is conditioned by subtracting a
  $5^\mathrm{th}$-order polynomial fit and then applying a
  normalization factor equal to the curve's standard deviation.
Vertical dotted lines in each plot
  lie, from left to right, at $\omega = \omegalump$, $\Omegabin$, and
  $2\left(\Omegabin - \omegalump \right)$; for those runs without a
  lump, $\omegalump$ of \origrun is used instead.  
  (Left-to-right) \massratiolist. 
\label{fig:light-mdot-curves-fft-mass-ratios}}
\end{figure*}

\begin{figure*}[htb]
\centerline{
\includegraphics[width=2\columnwidth]{\plotdir/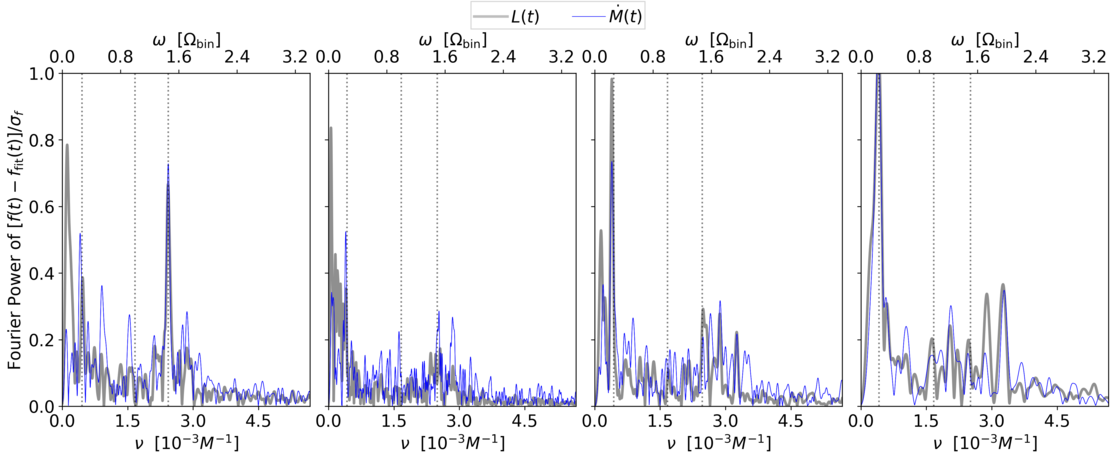}
}
\caption{Fourier power spectra of the light curves and accretion rates, $\dot{M}(r=a,t)$,
  including only times $t > \Tlump$.  For those runs with no
  observed lump, we use the simulation's last $2.5\times 10^4 \mbh$ of
  time.  Before performing the Fourier power spectrum, the function is
  conditioned by subtracting a $5^\mathrm{th}$-order polynomial fit
  and then applying a normalization factor equal to the curve's
  standard deviation.
Vertical dotted lines in each plot
  lie, from left to right, at $\omega = \omegalump$, $\Omegabin$, and
  $2\left(\Omegabin - \omegalump \right)$; for those runs without a
  lump, $\omegalump$ of \origrun is used instead.  
(Left-to-right)   \magfluxlist. 
\label{fig:light-mdot-curves-fft-init-states}}
\end{figure*}

\begin{figure*}[htb]
\centerline{
\includegraphics[width=2\columnwidth]{\plotdir/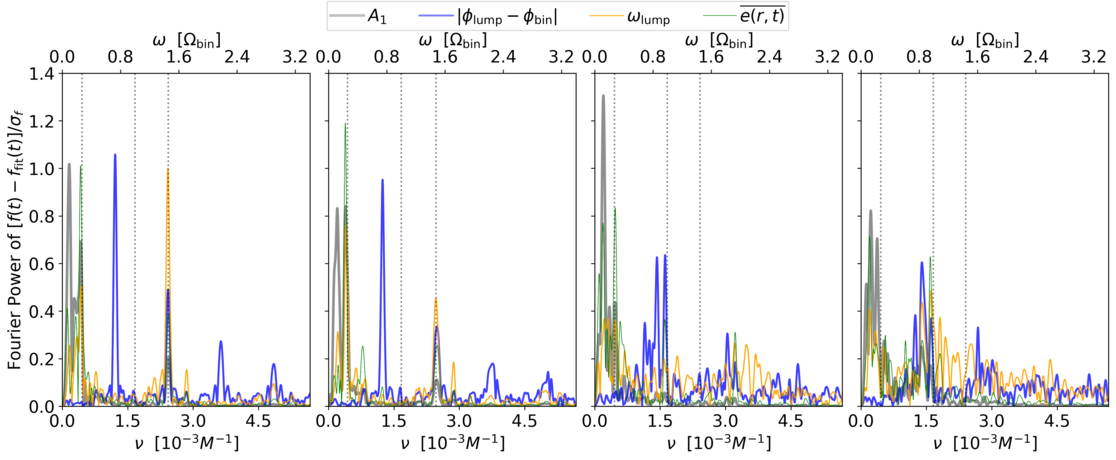}
}
\caption{Fourier power spectra of
  quantities related to the lump's amplitude,
  motion, and position
  including only times $t > \Tlump$. For those runs with no
  observed lump, we use the simulation's last $2.5\times 10^4 \mbh$ of
  time.  Before performing the Fourier power spectrum, the function is
  conditioned by subtracting a $5^\mathrm{th}$-order polynomial fit
  and then applying a normalization factor equal to the curve's
  standard deviation.
Vertical dotted lines in each plot
  lie, from left to right, at $\omega = \omegalump$, $\Omegabin$, and
  $2\left(\Omegabin - \omegalump \right)$; for those runs without a
  lump, $\omegalump$ of \origrun is used instead.  
  (Left-to-right) \massratiolist.
\label{fig:timeseries-all-mass-ratios}}
\end{figure*}

\begin{figure*}[htb]
\centerline{
\includegraphics[width=2\columnwidth]{\plotdir/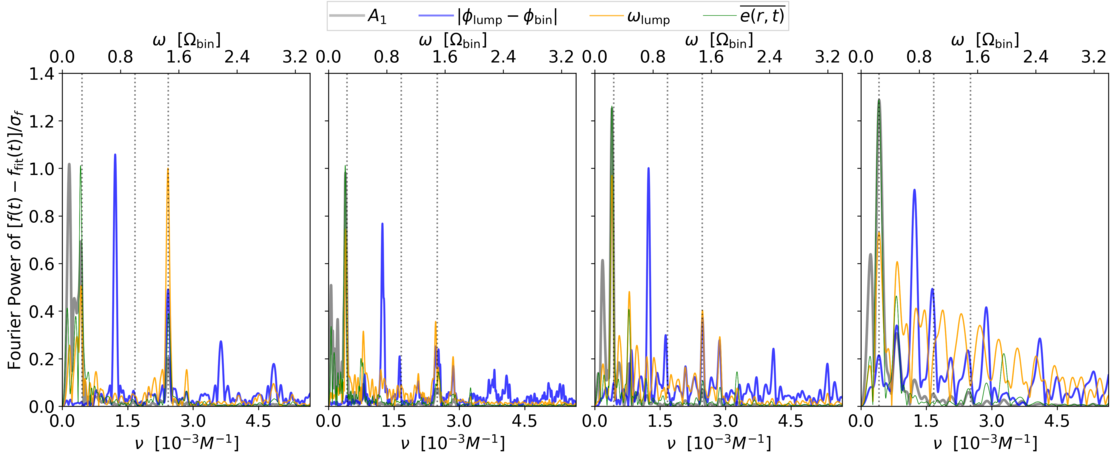}
}
\caption{
Fourier power spectra of
  quantities related to the lump's amplitude,
  motion, and position
  including only times $t > \Tlump$. For those runs with no
  observed lump, we use the simulation's last $2.5\times 10^4 \mbh$ of
  time.  Before performing the Fourier power spectrum, the function is
  conditioned by subtracting a $5^\mathrm{th}$-order polynomial fit
  and then applying a normalization factor equal to the curve's
  standard deviation.
Vertical dotted lines in each plot
  lie, from left to right, at $\omega = \omegalump$, $\Omegabin$, and
  $2\left(\Omegabin - \omegalump \right)$; for those runs without a
  lump, $\omegalump$ of \origrun is used instead.  
(Left-to-right)   \magfluxlist.
\label{fig:timeseries-all-init-states}}
\end{figure*}

In Figure~\ref{fig:light-mdot-curves-mass-ratios} we show the light
curve and accretion rate for each simulation in the
\massratioruns.   After the initial
burst of accretion, i.e.  $t > 2.5\times10^4\mbh$,  the
luminosity in each run generally tracks the run's accretion rate,
implying that the radiative efficiency, $\eta \equiv L / \dot{M}$,
remains nearly constant. Closer inspection, however,  shows a sharp increase in $\eta$ at the time of maximum
luminosity; after this time, $\eta$ declines slowly from this elevated value.  Similar trends are also seen in the \magfluxruns,
in Figure~\ref{fig:light-mdot-curves-init-states}.

In all cases,
the accretion rate exhibits larger relative amplitude and higher
frequency fluctuations than the luminosity does, an observation we
found with single-BH accretion disk simulations using similar
thermodynamics \citep{NK09}.
The luminosities of the \massratioruns
are all quite similar, demonstrating that the mass-inflow provided by the
outer disk is the dominant regulator of light output.
 
Although the magnitude of the luminosity in the \magfluxruns
is quite similar to that of the \massratioruns, their time-dependence is quite
different.   \medrun and \lrgrun both exhibit a plateau period in their
luminosities following the initial peak.
 The perturbation imparted in \injectrun is apparent in both
$L(t)$ and $\dot{M}(t)$.  In effect, it causes the disk to go through two
cycles of rise-and-fall, rather than the single one of the other simulations.
Interestingly, there is a short delay ($\sim 2000M$) between the second peak in accretion
and the subsequent peak in luminosity.

Phase alignment in the fluctuations of $L(t)$ and $\dot{M}(t)$ for the
\massratioruns and \magfluxruns are best seen through
their normalized correlations, which we show in 
Figure~\ref{fig:light-mdot-curves-corr-mass-ratios} and
Figure~\ref{fig:light-mdot-curves-corr-init-states}, respectively.
We calculate the normalized correlations between $L(t)$ and $\dot{M}(t)$ using
\beq{ \mathtt{Corr}\left[L, \dot{M}\right](\Delta t) = \frac{\sum_i
    L\left(t_i\right) \dot{M}\left(t_i-\Delta t\right) }{\lnorm{2}{L}
    \ \lnorm{2}{\dot{M}}} \quad ,
  \label{correlation-eq}  
}
where $\lnorm{2}{L}$ ($\lnorm{2}{\dot{M}}$) is the $l^2$-norm of the
luminosity (accretion rate) time series minus a $5^\mathrm{th}$-order
polynomial fit to the raw data to remove secular trends. For runs clearly
exhibiting a lump (\qone and \qtwo), the correlations show larger peak amplitudes than
for those without (\qfive and \qten).
In addition, the lump runs show variability on both the $\tlump$ ($\simeq 4\tbin$) and twice the
beat frequency ($1.5\Omegabin$) time scales, though \origrun shows
relatively weak lump period variability.  Because
lump-producing runs all show correlations that peak at positive
lags, the luminosity variation follows that of the
accretion rate; this reflects the sequence of events in which an accretion
stream leaves the inner edge of the circumbinary disk, part of it feels sufficient
torque to return to the circumbinary disk, and once it arrives, dissipates some
of its energy in a shock, whose heat is then radiated \citep{Noble12,Shi15}.

Using the detrended and normalized functions of time, we plot the
normalized Fourier power distributions in
Figure~\ref{fig:light-mdot-curves-fft-mass-ratios} and
Figure~\ref{fig:light-mdot-curves-fft-init-states} for the
\massratioruns and \magfluxruns, respectively. The power spectra are
calculated using the period $t>\Tlump$ if the run exhibits a lump, and
$t > \tend - 2.5\times10^4\mbh$ if not.

The character of the variability changes with mass ratio.  Neglecting the peak at
very low frequency, which could be an artifact of the detrending, the nature of
the strongest peak in the luminosity power spectrum is different in each case.
In \origrun, it is at $\simeq 1.5 \Omegabin$, twice the beat frequency between
the lump orbital frequency and the binary orbital frequency.
In \qtwo, there is also a strong peak at twice the beat frequency for ${\dot M}$,
but the peak for $L$ is much smaller than in \qone, and there is a comparable peak at 
$\simeq 0.75 \Omegabin$, the actual beat frequency.
In \qfive, there is a very strong peak at almost exactly $\Omegabin$.
Lastly, in \qten, there is no significant periodic behavior in the light
output at all. 
The presence of a small peak at $\Omegabin$ in \qtwo and a larger
one in \qfive may be interpreted as due to the closer approach of the
secondary to the inner edge of the circumbinary disk as $q$ decreases,
and the consequent enhancement of modulation at the secondary's orbital
frequency;
the disappearance of this peak in \qten is likely a sign that when the mass-ratio
is this small, the secondary has hardly any affect on the accretion.
In all cases, there are contrasts between the power spectra of the 
accretion rate and the luminosity; in other words, there are significant
contributions to the rate of heat dissipation that are {\it not} due immediately
to mass accretion.

In Table~\ref{tab:variability} we provide the relative standard
  deviations of fluctuations in the accretion rate and luminosities.
 The most striking feature is that the fractional variation in
  accretion rate is consistently an order of magnitude larger than
  the fractional variation in the luminosity.   This fact, too, strongly
  indicates that the luminosity is not directly related to the accretion
  flow.
  No clear trends exist in the relative variabilities within each
  series, though if a run is more variable than another in one
  quantity it typically is more variable in the other quantity as
  well.

  \begin{table}[h]
  \hspace{-1.4cm}
  \begin{tabular}{|lcc|cc|}
\hline
\textbf{Run Name} &\textbf{$\sigma_{\dot{M}}/\overline{\dot{M}}$} &\textbf{$\overline{\dot{M}} [10^{-3}]$} &\textbf{$\sigma_{L}/\overline{L}$} &\textbf{$\overline{L} [10^{-4}]$} \\
\hline 
\hline 
\origrun        &$0.29$ &$5.6$ &$0.027$ &$3.3$ \\
\qtwo           &$0.54$ &$3.3$ &$0.048$ &$1.9$ \\
\qfive          &$0.33$ &$2.2$ &$0.025$ &$1.4$ \\
\qten           &$0.20$ &$4.8$ &$0.022$ &$1.8$ \\
\medrun         &$0.30$ &$11.$ &$0.043$ &$5.4$ \\
\lrgrun         &$0.38$ &$10.$ &$0.033$ &$5.1$ \\
\injectrun      &$0.56$ &$4.1$ &$0.054$ &$1.4$ \\
\hline 
\end{tabular}
 \caption{Standard deviations $\sigma_{\dot{M}}$ ($\sigma_{L}$) of accretion rate (luminosity)
      for each run, taken over the same period
      in which the PSDs were calculated in Figures~\ref{fig:light-mdot-curves-fft-mass-ratios}~-~\ref{fig:light-mdot-curves-fft-init-states}.
      Each standard deviation is normalized by the mean of the quantity in question over this period.  These averages are also displayed, though in code units. } 
\label{tab:variability}
\end{table}

We plot the PSDs of different quantities related to the lump in
Figures~\ref{fig:timeseries-all-mass-ratios}~and~\ref{fig:timeseries-all-init-states}.
The functions analyzed are the $m=1$ mode amplitude of the density
integrated over the lump region ($\int_{2a}^{4a} A_1 dr$), the phase
difference between the binary and the lump's phase ($\left|
\phi_\mathrm{lump} - \phi_\mathrm{bin}\right|$), the orbital frequency
of the lump ($\omegalump$), and $\elump(t)$.  The phase of the lump is
found by locating the maximum of the $m=1$ mode amplitude integrated
over the radial extent of the lump region.
  
The lump's orbital frequency can then be defined by $\omegalump \equiv
\dot{\phi}_\mathrm{lump}$.  The average $\omegalump$ over this period
(for those runs with a lump) is reported in
Table~\ref{tab:run-characteristics}; it is consistently $0.26
\Omegabin$ or $\simeq \Omegabin/4$ for all runs, suggesting a 4:1
resonance between the lump and binary.  As expected, in the runs
exhibiting a lump, the phase difference is modulated at the beat
frequency $\left(\Omegabin-\omegalump\right)$ and its higher
harmonics.
 
Several aspects of this timing analysis confirm the coherence of lump
motion.  Plots of $\omegalump(t)$ are continuous.  Most clearly seen
in \qone and---to a lesser extent---in the other lump-forming runs,
$\omegalump$ varies at twice the beat frequency, suggesting that the
lump is accelerated by each passing BH.  The eccentricity of lump
orbits fluctuates primarily at the lump's average Keplerian rate,
$\omegalump$ as well as at twice the beat frequency; this, too,
indicates that the lump is a distinct physical element, not a pattern.

In order to connect the accretion rate and light curve variability to
the lump, we compare their PSDs (in
Figure~\ref{fig:light-mdot-curves-fft-mass-ratios}) to those associated
with the density structure of the lump-forming region (in
Figure~\ref{fig:timeseries-all-mass-ratios}).  Strikingly, in both sets
of PSDs, the most prominent peaks for \qone and \qtwo, the runs with
the clearest lumps, occur at the same frequencies, twice the beat
frequency and $\omegalump$ for $q=0.5,1$.

For the \magfluxruns we compare
Figure~\ref{fig:light-mdot-curves-fft-init-states} to
Figure~\ref{fig:timeseries-all-init-states}.  Relative to the other
runs, those exhibiting larger or smaller fluctuation power in the
lump's properties at $\omegalump$ and twice the beat frequency also
exhibit larger or smaller signals in $L(t)$ and $\dot{M}(t)$ at these
frequencies.  \injectrun, which is the only run of this series that
does not show significant variability in $\left|\phaselump -
\phasebin\right|$ at $2\left( \Omegabin - \omegalump\right)$, does
show significant variability in this quantity at twice the other beat
frequency, $2\left( \Omegabin + \omegalump\right)$ as does its
accretion rate.

\section{Discussion}
\label{sec:discussion}

\subsection{Origins and Conditions of Lump Formation}
\label{sec:orig-lump-accr}

A primary goal of our paper is to investigate how and why the lump
forms, and what conditions are amenable to the lump's growth.  We used
multiple diagnostics to verify the presence of a coherent, orbiting,
overdense region of gas. In particular, the surface density, cavity
wall eccentricity, lump phase and orbital velocity, and the PSD
spectra of several quantities all consistently show signs of the lump
when one is present.

The \massratioruns demonstrated that sufficiently large mass ratios are
required to manifest a significant lump.
From the \magfluxruns, we learned that the amount of mass in the disk had
little effect on the lump, but the amount of magnetic flux did: adding a
relatively modest, ordered, poloidal magnetic field distribution
beyond the lump region was enough to perturb it to the point the
nascent lump was disrupted and formed approximately $\simeq 100 \tbin$
or $25 \tlump$ later.  Also, the time, $\Tlump$, at which a run passed
our lump-formation criterion, increased with decreasing mass ratio
(gravitational torque) and with increasing available
mass/magnetic-flux. These results suggest that a circumbinary disk's
ability to form a lump is robust to minor deviations in conditions,
though requires a sufficiently strong gravitational torque.  

Although MHD turbulence usually has the most power on the longest spatial
wavelength modes, implying that small $m$ azimuthal  modes  have
the most power, a distinct mechanism for sustaining a \textit{coherent} $m=1$ mode is
required, as a $m=1$ turbulent mode would be \textit{incoherent}.
This coherence is supplied by the binary's gravitational torque in
two ways.   One has previously been cited: the lump is reinforced by those
portions of the accretion streams thrown back to the circumbinary disk by
the gravitational torque \cite{Shi12,DOrazio13}.   Phase coherence is further
maintained because the lump's orbit is resonant with the binary's
orbit: $\tlump:\tbin = 4:1$.
The importance of this resonance is evident in these two time scales'
prominence in the PSDs of the accretion rate, luminosity, eccentricity, and $m=1$ density mode.

As we saw in \injectrun a perturbation to the magnetic field was
sufficient to disturb the lump, so why is the inherent magnetic field
insufficient to shear apart a growing $m=1$ fluctuation?  After all,
one typically finds, in disks about single black holes, the magnetic
stress per unit mass is nearly uniform in azimuth, with incoherent
fluctuations having a fairly smooth inverse polynomial power spectrum
w.r.t. wavelength.  Obviously there is a competition between the
forcing and the local shear stress.

From measuring the MRI quality factors
(Appendix~\ref{app:mri-resolution}), we know that magnetic field per
unit enthalpy degrades within the overdensity region. So how does the
resonant interaction encourage mass growth over magnetic field growth
in the circumbinary disk? For there to be a physical origin for the
depletion of specific magnetic field strength in the lump, we need to
understand how the magnetic field is preferentially destroyed there.
The mechanism also needs to depend on the mass ratio since we find
that a significant lump forms for only sufficiently large $q$.
The answer comes from animations of magnetic field structure in the torqued streams
striking the circumbinary disk, which show that the magnetic field in these
streams is directed opposite to the field in the disk where the stream
arrives.
The collision of oppositely-oriented magnetic
field distribution with the inner cavity wall material leads to
large-scale reconnection and dissipation of the field into heat.
This process can therefore explain how the
magnetic field in the lump region decreases.

Local magnetic field may grow through local MHD instabilities
like the MRI and be replenished by field carried into the region by
inward fluid motion.  Our interest in exploring
these processes was the reason for plotting the magnetic stress per
unit mass, ${W^r}_\phi$ in
Figures~\ref{fig:maxwell-to-density-spacetime-mass-ratio}~-~\ref{fig:maxwell-to-density-spacetime-states}.
Lumps form only when  ${W^r}_\phi $ falls below $\simeq 10^{-4}$
in the region near the circumbinary disk's inner edge.
While this is just a correlation,
it is one that works for runs with different $\Tlump$, suggesting it is
not a simple function of the mass ratio or initial conditions.  In
order to explore why this value is important, let us compare the time
scales for magnetic field advection across the lump, $\dtlump$, and
the time scale over which the magnetic field is dissipated, $\tdiss$,
by compression of expelled streams with oppositely oriented magnetic
field.

Assuming time steadiness of the accretion flow and that
Maxwell stress accounts for the
majority of the total stress, one can show that far from the edge of
the disk:
\beq{
  {W^r}_\phi \simeq r \Omega_K(r) \massavg{ u^r } \quad
  , \label{stress-per-mass-approx}
}
where $\Omega_K$ is the local
Keplerian orbital rate, and $\massavg{u^r}$ is the accretion inflow
speed which can be used to estimate the time scale for advection of plasma
across the lump, $\dtadv$:
\begin{eqnarray}
  \dtadv & = & 
  \frac{\drlump}{\massavg{u^r}} = 
  \left. \frac{ \drlump \, r \, \Omega_K(r) }{ {W^r}_\phi}
  \right|_{r=\rlump} \label{dtadv0} \\
   & = & \left(\frac{\drlump}{ a } \right) \left(\frac{\rlump }{ a } \right) \left(\frac{a}{\mbh} \right)^2 \omegalump \mbh^2 \left( {W^r}_\phi \right)^{-1} \label{dtadv1} \nonumber \\
  & = & \left(\frac{\drlump}{ a } \right) \left(\frac{\rlump }{ a } \right)^{-\frac{1}{2}}  \left( 2\pi {W^r}_\phi \frac{a}{\mbh} \right)^{-1} \, \tbin \label{dtadv2} \\
  & \simeq & 5 \tbin \left(\frac{\drlump}{0.1 a} \right) \left(\frac{\rlump}{2.5 a} \right)^{\frac{1}{2}} \nonumber \\
  &&\times  \left(\frac{a}{20\mbh}\right)^{-1}  \left(\frac{{W^r}_\phi}{10^{-4}} \right)^{-1} , \label{dtadv3} 
\end{eqnarray}
where we have used the Newtonian rotation rates, 
$\omegalump = \mbh^{-1} \left(\rlump/\mbh\right)^{-3/2}  = \mbh^{-1} \left(\rlump/a\right)^{-3/2}  \left(a/\mbh\right)^{-3/2}$, and
$\tbin = 2 \pi \mbh \left(a/\mbh\right)^{3/2}$.
The average radial extent of the growing lump, $\drlump$, is
often found to be a fixed fraction of the binary separation,
$a$.   This fraction is generically small because
the $m=1$ overdensity  originates from an expelled accretion stream compressed by its shock against the cavity wall.
We estimate $\drlump \sim 0.1 a$ at the time the lump begins to form, which has been observed in a number of
simulations \cite{MM08,Noble12,Zilhao2015,Farris14a,Miranda2017}.

The dissipation time scale of magnetic field loss in the lump is the period between successive
BH-overdensity interactions, which occur at twice the beat frequency
$\Omega_\mathrm{diss} = 2\left(\Omegabin - \Omega_K(\rlump) \right)
\simeq \frac{3}{2} \Omegabin$: 
\beq{
  \tdiss = \frac{2 \pi}{
    \Omega_\mathrm{diss}}
  \simeq \frac{2}{3} \tbin \quad . \label{tdiss}
}
If one process occurs at a faster rate, it will eventually win out.  
The ratio of the two time scales, $Y$, is therefore useful:
\beqa{
  Y  & \equiv &  \frac{\tdiss}{\dtadv}  \label{Y-ratio0} \\
  & \simeq & 0.13 \left(\frac{\drlump}{0.1 a} \right)^{-1} \left(\frac{\rlump}{2.5 a} \right)^{-\frac{1}{2}} \nonumber \\
  && \times \left(\frac{a}{20 \mbh} \right) \left(\frac{{W^r}_\phi}{10^{-4}} \right) .
  \label{Y-ratio1}
}
We would expect a lump to develop once $Y < 1$, and may not otherwise
because the MRI operates on a $\tlump \gg \tbin$ time scale at the
location of lump.  For our parameters, we find this ratio implies a
lump will grow.  We found that the lump does not occur earlier in the
lump-forming evolutions because ${W^r}_\phi$ is an order of magnitude
larger, pushing $Y>1$.  When we inject magnitude field in \injectrun,
${W^r}_\phi$ grows by an order of magnitude resulting in $Y>1$ until
the specific magnetic stress returns to the $10^{-4}$ level and the
lump returns. We also note that once the lump begins to form, its
radial extent grows, which makes it more difficult to rejuvenate its
magnetic field through advective mixing because $Y \propto
\drlump^{-1}$.

Although this model does not explicitly depend on the mass ratio, the
qualitative picture does help us understand why it is more difficult
for binaries with smaller $q$ to form a lump.  
First, the magnitude of the binary's
time-dependent quadrupole moment decreases for smaller $q$.  As a result, the binary's gravitational
torque weakens, diminishing how much of the matter entering the gap is
thrown back at the disk to feed the lump.
This effect may also decrease the effectiveness of magnetic field
dissipation for smaller $q$ because a smaller amount of mass returned to
the disk carries a smaller amount of (oppositely-directed) magnetic field,
thereby increasing $W^r_\phi$.
This trend is bolstered by the fact that
$Y$ increases by a further factor of 2 as the BH-lump interaction frequency falls
from twice the beat frequency (for $q=1$) to exactly the beat frequency (for $q \ll 1$).

Simulations employing viscous hydrodynamics have also
 demonstrated lumps
\citep{MM08,DOrazio13,Farris14a,Farris14b,DOrazio2016,Munoz2016,Miranda2017,Munoz2019,Moody2019,Mosta2019,Duffell2020,Zrake2020,Munoz2020a,Munoz2020b,Tiede2020},
 even though there can be no limitation of internal stresses due to magnetic
reconnection in them.   It is possible that the nature of their viscous model for
internal stress serves the same purpose because the lump's coherent motion
eliminates internal differential rotation, thereby reducing the internal viscous stress.

\subsection{Comparison to Prior Work}

In the following, we will summarize how our results
  compare and contrast with prior work, limiting our comparison to those studies using
  binaries on fixed circular orbits with prograde disks.

By all accounts, the lump is a significant density enhancement that forms near or along the cavity wall,    
spans a relatively narrow radial extent, and orbits as a coherent
structure at the local Keplerian rate (Section~\ref{sec:non-axisymm-struct} and 
\cite{MM08,DOrazio13,Farris12,Shi12,Noble12,Gold14,Zilhao2015,DOrazio2016,Bowen2018,Bowen2019,LopezArmengol2021,Paschalidis2021}).
 In its asymptotic state,
$\Delta \rlump \simeq a$, while spanning a significant fraction of the
possible azimuthal extent $\pi/3 \lesssim \delta\phi_\mathrm{lump}
\lesssim \pi$.  VH simulations deviate a little from this picture, at least
for the larger $q$ cases that exhibit significant $m=1$ structure, in that
they also develop a density enhancement at the apoapse
\citep{MM08,DOrazio13,Farris14b,DOrazio2016,Miranda2017,Ragusa2020}
that precesses at a much slower rate than the local Keplerian velocity
of the disk. Some  report that the lump travels through this ``traffic jam'' of gas
  at the local Keplerian rate \citep{Ragusa2020}, 
while others show the lump
vanish and return cyclicly \cite{Miranda2017}.

Regardless of differences in the azimuthal distribution of the lump,
all\footnote{Few numerical relativity simulations have the temporal
range to calculate a PSD of $\dot{M}(t)$, and those that have PSDs
that have relatively poor frequency resolution and dynamic range to
yield significant signal-to-noise \citep{Gold14}.}  simulations find
that it leads to accretion rate modulation at a time scale
approximately equal to the Keplerian period at cavity's edge.  This
modulation time scale usually is approximately $\tlump \simeq 4-5
\tbin$ for cases with $h/r \sim 0.1$
\citep{MM08,Shi12,Noble12,DOrazio13,Farris14b,Zilhao2015,DOrazio2016,Munoz2016,Miranda2017,Moody2019,Munoz2020a,Duffell2020,LopezArmengol2021,DittmannRyan2021}. All
our simulations with a lump show $\tlump \simeq 4 \tbin$, which were
the cases when $q>0.2$.  Interesting,
\cite{Duffell2020,DittmannRyan2021} also report the modulation changes
rather sharply at $q=0.2$ for circular orbits and $h/r=0.1$ disks; and
they both found that $\tlump$ gradually decreased from $5\tbin$ at
$q=1$ to $4\tbin$ at $q\rightarrow0.2$.  Others show transitions in
the same ballpark: $q>0.25-0.5$ \citep{DOrazio13}, $q>0.25-0.43$
\citep{Farris14b}, $q\ge0.4$ \citep{Munoz2020a}.  For cases with
larger cavities (e.g., with colder disks), longer periods are found:
\cite{Ragusa2016} find $\tlump \simeq 5\tbin$ ($\tlump \simeq 4\tbin$)
for $h/r=0.04$ ($h/r=0.1$), and \cite{Ragusa2020} report modulations
at $7-8\tbin$ from a cavity of radius $r\simeq 3.5a$.
A dissenting
case is the $q=0.1$ simulation of \cite{Shi15} which shows
evidence of the lump in the surface density and the accretion rate,
with the lump even forming later in their $q=1$ case; the difference
may be due to their use of an isothermal equation of state.
As our  MHD simulations, like those of others, include turbulent
  circumbinary disks, our PSDs of $\dot{M}(t)$ are more complex than those of VH simulations
  and include significant 
  power over a range of frequencies, like those observed in real AGN disks.  Demonstrating that
  variability associated with the lump is evident above these broadband fluctuations for
  several different cases is a key finding of this paper.

Few papers, however, investigate how the lump forms and endures.
\cite{Shi12,DOrazio13,Shi15} went to significant lengths to show
how returning streams reinforce the lump.  In \cite{Noble12} we showed
how the MRI quality of our simulation degraded in the lump; this
analysis is repeated for our simulations here in
Appendix~\ref{app:mri-resolution}.  Evidence of degradation of
magnetic and hydrodynamic turbulence in the lump is shown in
\cite{Shi2016} by spatially associating the lump with the relative
contributions of the vertical components of the velocity and magnetic
field.  And, \cite{Ragusa2017} explore whether the lump is stabilized
by vorticity, which they found to not be the case as vorticity was
shown to be smaller in the lump than in the rest of the circumbinary
disk.  All evidence presented so far is consistent with the notion
that the lump is a coherently orbiting structure with a deficit or
absence of internal differential rotation or local vorticity; as we
have mentioned, the lack of differential rotation helps to stabilize
it by hindering the MRI or viscous stresses.

\section{Summary}
\label{sec:conclusions}

We have explored two series of simulations of circumbinary accretion
disks about binary black holes using GRMHD and 2.5PN 
approximate spacetimes: the \massratioruns, in which the relative masses
of the black holes were $q=0.1, 0.2, 0.5,$ or  1, while all other aspects
remained the same, and the \magfluxruns, which used an equal mass
binary but different distributions of mass and magnetic field.
Axisymmetric and non-axisymmetric properties of the circumbinary disks
were explored in each series.  Special emphasis was given to aspects of the
circumbinary disk that may affect binary signatures in EM emission,
such as the lump.

In terms of axisymmetric aspects, we found that the circumbinary disk
in most cases approached a steady state of mass inflow by the end of
each run.  Aspects such as the surface density peak (gradient) were
found to broaden (move inward) as the mass ratio decreased.  Runs with
larger distributions of mass/magnetic-flux took longer to asymptote to
a steady state, but otherwise resembled the other equal-mass runs in
axisymmetric aspects.  

Regarding the non-axisymmetric properties, we explored the disk's
azimuthal structure, eccentricity, and variability. We found a new
diagnostic for the relative strength of the $m=1$ azimuthal mode of
density such that lump formation was always sustained if this quantity
exceeded a certain value.  The growth in time of this diagnostic was
found to be coincident with the growth rate of eccentricity of
material near the cavity.  When arising, the lump is found to be a
coherent structure in which magnetic stress is kept
comparatively weak by repeated collisions with expelled material
carrying oppositely-oriented magnetic field. The lump is also
associated with variability in the light curve and accretion rate on
two time scales related, {\it but not identical, to} the binary period:
$4\tbin$ and $2/3 \tbin$, the former associated with the lump's local
Keplerian period, the latter with the beat mode between the binary and
the lump.
 
The lump was observed in $q=0.5, 1$ and not found in $q=0.2,0.1$,
suggesting the no-lump/lump transition lies between $0.2<q<0.5$ for
our simulation conditions.  We also demonstrated that persistent
reservoirs of mass and magnetic flux available to be accreted into the
lump region do not hinder its growth.  However, a perturbation to the
accretion flow in the form of a modest additional ordered magnetic
field was sufficient to disrupt a nascent lump and delay its
development for $\sim 10\tbin$.

Previous work on the origin and sustenance of lumps stressed the
coordinated delivery of matter to the lump region by ``torqued-up''
streams  \citep{Shi12,DOrazio13,Shi15}.  We have found that
the matter returned to the disk in this
way carries with it magnetic field that tends to be directed opposite
to the already-existing field in the lump.  Reconnective dissipation
then suppresses the magnetic stress per unit mass in the lump, helping
the lump region retain its mass.  As a result, lumps grow once the
magnetic stress per unit mass falls below a critical level, a
threshold that is essentially independent of simulation parameters.
  
Because we have not explored the entire parameter space of conditions,
our conclusions are limited to the choices made herein (e.g., the
thermodynamic model, disk aspect ratio).  We will reserve exploring
other parts of parameter space to future work.  Further, these new,
relaxed circumbinary disk simulations provide starting conditions with
which we may pursue simulations with resolved black holes
\citep{Bowen2018,dAscoli2018,Bowen2019}, as we have done using
circumbinary disk data of \cite{Noble12}


\begin{acknowledgments}

We thank Carlos Lousto (RIT), Federico Lopez Armengol (RIT), Luciano Combi
(Instituto Argentino de Radioastronom\'ia), Mark Avara (Cambridge),
and Dan D'Orazio (NBI, Copenhagen) for valuable discussions on work
described in this manuscript.

This work was supported by several National
Science Foundation (NSF) grants. 
S.C.N., M.C., M.Z., and Y.Z. received support from awards AST-1028087 
and J.H.K. from AST-1028111, PHY-1707826, and AST-2009260.
M.C., B.C.M, H.N., Y.Z. also acknowledge partial support 
from grants PHY-0929114, PHY-0903782, PHY-0969855, and PHY-1707946.
Addtionally, M.C  was supported by awards AST-1516150 and AST-2009330, and 
S.C.N.  was supported by awards AST-1515982 and
OAC-1515969.

H.N. acknowledges support from JSPS KAKENHI Grant Nos. JP16K05347 and JP17H06358.

S.C.N. was supported by an appointment to the NASA Postdoctoral
Program at the Goddard Space Flight Center administrated by USRA
through a contract with NASA.

This research is part of the Blue Waters sustained-petascale computing 
project, which is supported by the NSF 
(awards OCI-0725070 and ACI-1238993) and the state of Illinois. Blue Waters 
is a joint effort of the University of Illinois at Urbana-Champaign and 
its National Center for Supercomputing Applications.  This work is also 
part of the ``Computational Relativity and Gravitation at Petascale: 
Simulating and Visualizing Astrophysically Realistic Compact Binaries'' 
PRAC allocation support by the NSF (award
OCI 0832606).  All but one simulation ($q=1/2$) were performed 
on Blue Waters.  The remaining 
simulation was performed on Stampede at the Texas Advance Computing Center
through the XSEDE allocation TG-PHY060027N.  Analysis was performed 
on the  NewHorizons and BlueSky Clusters at Rochester
Institute of Technology, which were supported by NSF grant No. PHY-0722703,
DMS-0820923, AST-1028087, and PHY-1229173.

This work was performed in part at Aspen Center for Physics, which is
supported by NSF grant PHY-1607611.

This research was supported by the Munich Institute for Astro- and
Particle Physics (MIAPP) which is funded by the Deutsche
Forschungsgemeinschaft (DFG, German Research Foundation) under
Germany's Excellence Strategy – EXC-2094 – 390783311.

\end{acknowledgments}

\software{astropy \citep{astropy2013},
  \harm \citep{GMT03,Noble06,Noble09,Noble12} }

\newpage
\appendix


\section{NZ Metric Range of Validity}
\label{app:NZValidity}

In order to discuss the range of validity for the NZ metric, we must first discuss the two types of PN approximation.
The first is the PN approximation to the metric.
For simplicity, we consider $g_{rr}=1/(1-2M/r)$ of the Schwarzschild metric 
in the Schwarzschild coordinates, and the series expansion with respect to $M/r$ up to 2PN order,
$g_{rr}^{\rm (3PN)} = 1 + 2M/r + 4M^2/r^2$.
In practice, we do not have $O((M/r)^3)$ terms in the spatial component of the NZ metric.
The error due to the PN truncation is evaluated as
$|g_{rr}^{\rm (2PN)}/g_{rr}-1|=1.6\%$, $0.8\%$ and $0.24\%$ for $r/M=8$, $10$ and $15$, respectively.

The second type of PN approximation is in the PN equations of
motion. Here, we apply the result shown in
\cite{Sago:2016xsp} and \cite{Fujita:2017wjq} which are  extensions of
\cite{Yunes:2008tw} and \cite{Zhang:2011vha} for the PN region of validity for
quasi-circular orbits of a point particle orbiting around a massive
BH.  According to the appendix of \cite{Sago:2016xsp}, we have the
radius of convergence for the orbital velocity $v$ around $v \approx
0.5$.  Considering the orbital velocity as $v=\sqrt{M/r}$, the above
fact means that we may use $1/(1-4M/r)$ as a resummed form. The
series expansion with respect to $M/r$ up to 3PN order that is used in
this paper, becomes $1 + 4M/r + 16M^2/r^2 + 64M^3/r^3$. From the
similar analysis to the metric case, we have the PN truncation error,
$6.3\%$, $2.6\%$ and $0.51\%$ for $r/M=8$, $10$ and $15$,
respectively.  The above two analyses give only the PN truncation
errors, and in practice, we need to compare the PN orbital evolutions
with those obtained by numerical relativity simulations for BBHs (see,
e.g., \cite{Szilagyi:2015rwa} for comparisons of the waveforms).
According to \cite{Ajith:2012az} which gives a guideline for a
gravitational-wave frequency to hybridize the PN and numerical
relativity waveforms, the PN approximation may be good up to $r/M \sim
9$.  Furthermore, from the right figure of Figure~3 in
\cite{Mundim:2013vca}, which is shown for the $M_1=M_2=M/2$ and
$a/M=20$ case, the accuracy of the NZ and inner zone metrics becomes
comparable around $r/M_1=5$, i.e., $r/M=10$.  From the above
observations, the NZ metric is appropriate for the region of both $r_1
\gtrsim 10M_1$ and $r_2 \gtrsim 10M_2$ to describe the spacetime, and
this gives ${\rin}/{a} \gtrsim {(1/2 + q)}/{(1+q)}$ from the primary
BH and ${\rin}/{a} \gtrsim {(1 + q/2)}/{(1+q)}$ from the secondary BH.
Therefore, the constraint on $\rin$ is derived as
Eq.~\eqref{eq:rin_vs_a}.


\section{MRI Resolution}
\label{app:mri-resolution} 
The ability for a finite volume/difference code to adequately resolve
MHD turbulence depends largely on whether the fastest growing mode of
the MRI is resolved \cite{2004ApJ...605..321S}.  By performing a
series of resolution studies of global MHD disks,
\cite{hgk11,2012ApJ...749..189S} found that global, extrinsic
characteristics of the accretion flow asymptote with resolution.  They
found that those simulations that met or surpassed a particular set of
resolution criteria would reside in the asymptotic regime.  We follow
these guidelines in constructing the simulations here, just as we did
in \cite{Noble12} and use the same resolution before. Please refer to
Appendix~B of \cite{Noble12} for more details.

\begin{figure*}[htb]
  \includegraphics[width=0.71\columnwidth]{\plotdir/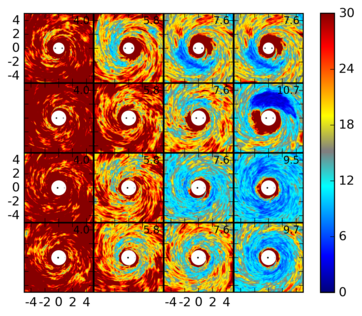}
  \includegraphics[width=0.71\columnwidth]{\plotdir/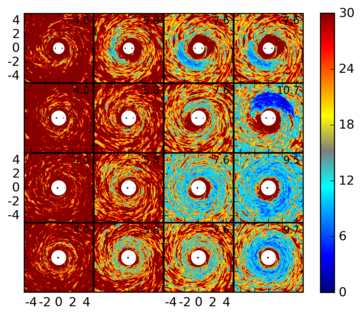}
  \includegraphics[width=0.71\columnwidth]{\plotdir/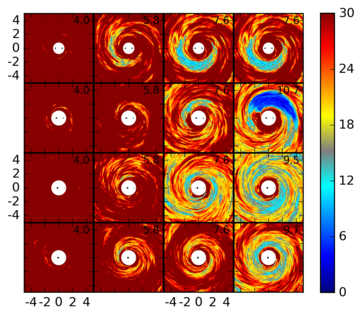}
\caption{(Left to right) The mass-weighted vertically integrated MRI
  quality factor, $\langle \QQ{i} \rangle_{\rho}$, in the radial,
  polar, and azimuthal directions, respectively, at three overlapping
  times and the final time of each simulation.  The times of each
  snapshot are specified in the upper-right corner of each frame in
  units of $10^4 \mbh$.  The vertical and horizontal axes are in units
  of $a=20\mbh$.  (Top to bottom) \qone, \qtwo, \qfive, \qten.
\label{fig:mri-q1-mass-ratios}}
\end{figure*}

\begin{figure*}[htb]
  \includegraphics[width=0.71\columnwidth]{\plotdir/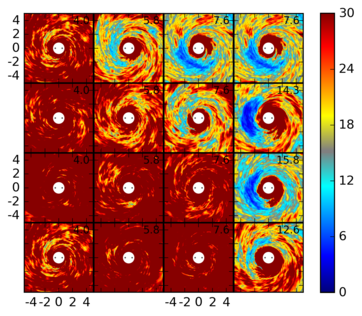}
  \includegraphics[width=0.71\columnwidth]{\plotdir/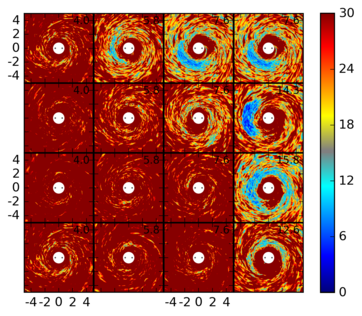}
  \includegraphics[width=0.71\columnwidth]{\plotdir/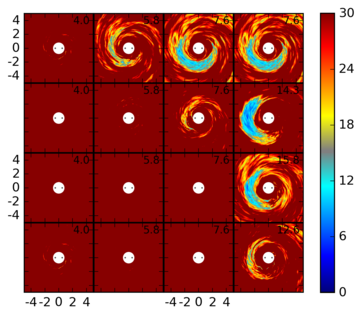}
\caption{(Left to right) The mass-weighted vertically integrated MRI
  quality factor, $\langle \QQ{i} \rangle_{\rho}$, in the radial,
  polar, and azimuthal directions, respectively, at three overlapping
  times and the final time of each simulation.  The times of each
  snapshot are specified in the upper-right corner of each frame in
  units of $10^4 \mbh$.  The vertical and horizontal axes are in units
  of $a=20\mbh$.  (Top to bottom) \origrun, \medrun, \lrgrun,
  \injectrun.
\label{fig:mri-q1-init-data}}
\end{figure*}

The MRI quality factors $\QQ{i}$ are ratios of the local MRI
wavelength and the local grid scale in a particular direction
\cite{Noble10}: \beq{ \QQ{i} = \frac{2 \pi \left|b^i\right|}{ \Delta
    \xx{i} \, \Omega_K(r) \, \sqrt{\rho h + 2 p_{m} } } \quad
  , \label{mri-quality-factors} } where index ``$i$'' denotes the
spatial numerical coordinate, with $i=1,2,3$ representing the radial,
poloidal, and azimuthal directions, respectively.  Averages in polar
angle ($\xtwo$) are mass-weighted  to bias the
integral over the turbulent portion near the disk's midplane rather
than the laminar regions of the corona and funnel: \beq{ \QQa{i} \equiv \frac{\int_0^{1} \QQ{i} \rho \,
    \sqrt{-g} \, d\xtwo}{ \int_0^{1} \rho \, \sqrt{-g} \, d\xtwo}
  \quad . \label{Qi-avg-x2} } The vertically-averaged quality factors,
$\QQa{1}$, $\QQa{2}$, $\QQa{3}$ are shown in
Figure~\ref{fig:mri-q1-mass-ratios} for the \massratioruns and
Figure~\ref{fig:mri-q1-init-data} for the \magfluxruns.  The quality factors
$\QQ{2}$ and $\QQ{3}$  are typically the most challenging to achieve, especially for
thinner disks.  This is why \cite{hgk11} recommends targets for
$\QQ{2}$ and $\QQ{3}$ of 10 and 25, respectively, above which the they
found simulations to be in the convergent regime.

In the \massratioruns (Figure~\ref{fig:mri-q1-mass-ratios}), we find
that all runs satisfy the $\QQ{2}>10$ and $\QQ{3}>25$ MRI quality
conditions for all times, with some exceptions.  The first exception
is that \qfive's quality factors diminish to just below their
threshold values for $t \gtrsim 7.6\times 10^4 \mbh$; \qten also shows
a decrease in MRI quality over this period, but not as steep a decline
in time, and only in localized regions are sub-threshold levels
reached.  Also, all quality factors are less than their targets in the
lump once it forms, as we found in \cite{Noble12}, too.

The \magfluxruns (Figure~\ref{fig:mri-q1-init-data}) on the other hand
is much better resolved in terms of its  MRI quality factors. The
only place where the MRI quality factors drop slightly below their
target values is within the lump toward the end of each run, with the
quality factors far exceeding the targets everywhere else.
Interestingly, the perturbation added in \injectrun seems to have lead
to conditions in which the quality factors are sustained at much
higher levels than otherwise, i.e. in \origrun; it is somewhat not
surprising that ordered poloidal field fosters more active MRI-driven turbulence, 
which is precisely why we added it in the first place.

In general, we consider our simulations well resolved, with the loss
of MRI quality explained as a natural consequence of binary-stream
interactions (Section~\ref{sec:orig-lump-accr}) or marginal enough to
not affect our qualitative conclusions.

\bibliography{bhm_references,references}

\begin{thebibliography}{}
\expandafter\ifx\csname natexlab\endcsname\relax\def\natexlab#1{#1}\fi
\providecommand{\url}[1]{\href{#1}{#1}}
\providecommand{\dodoi}[1]{doi:~\href{http://doi.org/#1}{\nolinkurl{#1}}}
\providecommand{\doeprint}[1]{\href{http://ascl.net/#1}{\nolinkurl{http://ascl.net/#1}}}
\providecommand{\doarXiv}[1]{\href{https://arxiv.org/abs/#1}{\nolinkurl{https://arxiv.org/abs/#1}}}

\bibitem[{Ajith {et~al.}(2012)}]{Ajith:2012az}
Ajith, P., {et~al.} 2012, Class. Quant. Grav., 29, 124001,
  \dodoi{10.1088/0264-9381/29/12/124001}

\bibitem[{{Astropy Collaboration} {et~al.}(2013){Astropy Collaboration},
  {Robitaille}, {Tollerud}, {Greenfield}, {Droettboom}, {Bray}, {Aldcroft},
  {Davis}, {Ginsburg}, {Price-Whelan}, {Kerzendorf}, {Conley}, {Crighton},
  {Barbary}, {Muna}, {Ferguson}, {Grollier}, {Parikh}, {Nair}, {Unther},
  {Deil}, {Woillez}, {Conseil}, {Kramer}, {Turner}, {Singer}, {Fox}, {Weaver},
  {Zabalza}, {Edwards}, {Azalee Bostroem}, {Burke}, {Casey}, {Crawford},
  {Dencheva}, {Ely}, {Jenness}, {Labrie}, {Lim}, {Pierfederici}, {Pontzen},
  {Ptak}, {Refsdal}, {Servillat}, \& {Streicher}}]{astropy2013}
{Astropy Collaboration}, {Robitaille}, T.~P., {Tollerud}, E.~J., {et~al.} 2013,
  \aap, 558, A33, \dodoi{10.1051/0004-6361/201322068}

\bibitem[{{Baker} {et~al.}(2019){Baker}, {Haiman}, {Rossi}, {Berger}, {Brandt},
  {Breedt}, {Breivik}, {Charisi}, {Derdzinski}, {D'Orazio}, {Ford}, {Greene},
  {Hill}, {Holley-Bockelmann}, {Key}, {Kocsis}, {Kupfer}, {Madau}, {Marsh},
  {McKernan}, {McWilliams}, {Natarajan}, {Nissanke}, {Noble}, {Phinney},
  {Ramsay}, {Schnittman}, {Sesana}, {Shoemaker}, {Stone}, {Toonen},
  {Trakhtenbrot}, {Vikhlinin}, \& {Volonteri}}]{Baker2019}
{Baker}, J., {Haiman}, Z., {Rossi}, E.~M., {et~al.} 2019, \baas, 51, 123.
\newblock \doarXiv{1903.04417}

\bibitem[{{Balbus} \& {Hawley}(1998)}]{BH98}
{Balbus}, S.~A., \& {Hawley}, J.~F. 1998, Reviews of Modern Physics, 70, 1

\bibitem[{{Bankert} {et~al.}(2015){Bankert}, {Krolik}, \& {Shi}}]{Bankert2015}
{Bankert}, J., {Krolik}, J.~H., \& {Shi}, J. 2015, \apj, 801, 114,
  \dodoi{10.1088/0004-637X/801/2/114}

\bibitem[{{Beckwith} {et~al.}(2008){Beckwith}, {Hawley}, \& {Krolik}}]{BHK08}
{Beckwith}, K., {Hawley}, J.~F., \& {Krolik}, J.~H. 2008, \apj, 678, 1180,
  \dodoi{10.1086/533492}

\bibitem[{Blanchet {et~al.}(1998)Blanchet, Faye, \& Ponsot}]{Blanchet:1998vx}
Blanchet, L., Faye, G., \& Ponsot, B. 1998, Phys. Rev., D58, 124002,
  \dodoi{10.1103/PhysRevD.58.124002}

\bibitem[{{Bowen} {et~al.}(2018){Bowen}, {Mewes}, {Campanelli}, {Noble},
  {Krolik}, \& {Zilh{\~a}o}}]{Bowen2018}
{Bowen}, D.~B., {Mewes}, V., {Campanelli}, M., {et~al.} 2018, \apjl, 853, L17,
  \dodoi{10.3847/2041-8213/aaa756}

\bibitem[{{Bowen} {et~al.}(2019){Bowen}, {Mewes}, {Noble}, {Avara},
  {Campanelli}, \& {Krolik}}]{Bowen2019}
{Bowen}, D.~B., {Mewes}, V., {Noble}, S.~C., {et~al.} 2019, \apj, 879, 76,
  \dodoi{10.3847/1538-4357/ab2453}

\bibitem[{{Charisi} {et~al.}(2016){Charisi}, {Bartos}, {Haiman},
  {Price-Whelan}, {Graham}, {Bellm}, {Laher}, \& {M{\'a}rka}}]{Charisi2016}
{Charisi}, M., {Bartos}, I., {Haiman}, Z., {et~al.} 2016, \mnras, 463, 2145,
  \dodoi{10.1093/mnras/stw1838}

\bibitem[{{Chen} {et~al.}(2020){Chen}, {Liu}, {Liao}, {Holgado}, {Guo},
  {Gruendl}, {Morganson}, {Shen}, {Zhang}, {Abbott}, {Aguena}, {Allam},
  {Avila}, {Bertin}, {Bhargava}, {Brooks}, {Burke}, {Carnero Rosell},
  {Carollo}, {Carrasco Kind}, {Carretero}, {Costanzi}, {da Costa}, {Davis}, {De
  Vicente}, {Desai}, {Diehl}, {Doel}, {Everett}, {Flaugher}, {Friedel},
  {Frieman}, {Garc{\'\i}a-Bellido}, {Gaztanaga}, {Glazebrook}, {Gruen},
  {Gutierrez}, {Hinton}, {Hollowood}, {James}, {Kim}, {Kuehn}, {Kuropatkin},
  {Lewis}, {Lidman}, {Lima}, {Maia}, {March}, {Marshall}, {Menanteau},
  {Miquel}, {Palmese}, {Paz-Chinch{\'o}n}, {Plazas}, {Sanchez}, {Schubnell},
  {Serrano}, {Sevilla-Noarbe}, {Smith}, {Suchyta}, {Swanson}, {Tarle},
  {Tucker}, {Norbert Varga}, \& {Walker}}]{Chen2020}
{Chen}, Y.-C., {Liu}, X., {Liao}, W.-T., {et~al.} 2020, \mnras, 499, 2245,
  \dodoi{10.1093/mnras/staa2957}

\bibitem[{{d'Ascoli} {et~al.}(2018){d'Ascoli}, {Noble}, {Bowen}, {Campanelli},
  {Krolik}, \& {Mewes}}]{dAscoli2018}
{d'Ascoli}, S., {Noble}, S.~C., {Bowen}, D.~B., {et~al.} 2018, \apj, 865, 140,
  \dodoi{10.3847/1538-4357/aad8b4}

\bibitem[{{Derdzinski} {et~al.}(2021){Derdzinski}, {D'Orazio}, {Duffell},
  {Haiman}, \& {MacFadyen}}]{Derdzinski2021}
{Derdzinski}, A., {D'Orazio}, D., {Duffell}, P., {Haiman}, Z., \& {MacFadyen},
  A. 2021, \mnras, 501, 3540, \dodoi{10.1093/mnras/staa3976}

\bibitem[{{Derdzinski} {et~al.}(2019){Derdzinski}, {D'Orazio}, {Duffell},
  {Haiman}, \& {MacFadyen}}]{Derdzinski2019}
{Derdzinski}, A.~M., {D'Orazio}, D., {Duffell}, P., {Haiman}, Z., \&
  {MacFadyen}, A. 2019, \mnras, 486, 2754, \dodoi{10.1093/mnras/stz1026}

\bibitem[{{Dittmann} \& {Ryan}(2021)}]{DittmannRyan2021}
{Dittmann}, A., \& {Ryan}, G. 2021, arXiv e-prints, arXiv:2102.05684.
\newblock \doarXiv{2102.05684}

\bibitem[{{D'Orazio} {et~al.}(2016){D'Orazio}, {Haiman}, {Duffell},
  {MacFadyen}, \& {Farris}}]{DOrazio2016}
{D'Orazio}, D.~J., {Haiman}, Z., {Duffell}, P., {MacFadyen}, A., \& {Farris},
  B. 2016, \mnras, 459, 2379, \dodoi{10.1093/mnras/stw792}

\bibitem[{{D'Orazio} {et~al.}(2013){D'Orazio}, {Haiman}, \&
  {MacFadyen}}]{DOrazio13}
{D'Orazio}, D.~J., {Haiman}, Z., \& {MacFadyen}, A. 2013, \mnras, 436, 2997,
  \dodoi{10.1093/mnras/stt1787}

\bibitem[{{Duffell} {et~al.}(2020){Duffell}, {D'Orazio}, {Derdzinski},
  {Haiman}, {MacFadyen}, {Rosen}, \& {Zrake}}]{Duffell2020}
{Duffell}, P.~C., {D'Orazio}, D., {Derdzinski}, A., {et~al.} 2020, \apj, 901,
  25, \dodoi{10.3847/1538-4357/abab95}

\bibitem[{{Farris} {et~al.}(2014){Farris}, {Duffell}, {MacFadyen}, \&
  {Haiman}}]{Farris14a}
{Farris}, B.~D., {Duffell}, P., {MacFadyen}, A.~I., \& {Haiman}, Z. 2014, \apj,
  783, 134, \dodoi{10.1088/0004-637X/783/2/134}

\bibitem[{{Farris} {et~al.}(2015){Farris}, {Duffell}, {MacFadyen}, \&
  {Haiman}}]{Farris14b}
---. 2015, \mnras, 446, L36, \dodoi{10.1093/mnrasl/slu160}

\bibitem[{{Farris} {et~al.}(2012){Farris}, {Gold}, {Paschalidis}, {Etienne}, \&
  {Shapiro}}]{Farris12}
{Farris}, B.~D., {Gold}, R., {Paschalidis}, V., {Etienne}, Z.~B., \& {Shapiro},
  S.~L. 2012, Physical Review Letters, 109, 221102,
  \dodoi{10.1103/PhysRevLett.109.221102}

\bibitem[{{Fontecilla} {et~al.}(2020){Fontecilla}, {Lodato}, \&
  {Cuadra}}]{Fontecilla2020}
{Fontecilla}, C., {Lodato}, G., \& {Cuadra}, J. 2020, \mnras, 499, 2836,
  \dodoi{10.1093/mnras/staa3071}

\bibitem[{Fujita {et~al.}(2018)Fujita, Sago, \& Nakano}]{Fujita:2017wjq}
Fujita, R., Sago, N., \& Nakano, H. 2018, Class. Quant. Grav., 35, 027001,
  \dodoi{10.1088/1361-6382/aa9ad5}

\bibitem[{{Gammie}(1996)}]{1996ApJ...457..355G}
{Gammie}, C.~F. 1996, \apj, 457, 355, \dodoi{10.1086/176735}

\bibitem[{{Gammie} {et~al.}(2003){Gammie}, {McKinney}, \& {T{\'o}th}}]{GMT03}
{Gammie}, C.~F., {McKinney}, J.~C., \& {T{\'o}th}, G. 2003, \apj, 589, 444,
  \dodoi{10.1086/374594}

\bibitem[{{Gold}(2019)}]{Gold2019}
{Gold}. 2019, Galaxies, 7, 63, \dodoi{10.3390/galaxies7020063}

\bibitem[{{Gold} {et~al.}(2014{\natexlab{a}}){Gold}, {Paschalidis}, {Etienne},
  {Shapiro}, \& {Pfeiffer}}]{Gold14}
{Gold}, R., {Paschalidis}, V., {Etienne}, Z.~B., {Shapiro}, S.~L., \&
  {Pfeiffer}, H.~P. 2014{\natexlab{a}}, \prd, 89, 064060,
  \dodoi{10.1103/PhysRevD.89.064060}

\bibitem[{{Gold} {et~al.}(2014{\natexlab{b}}){Gold}, {Paschalidis}, {Ruiz},
  {Shapiro}, {Etienne}, \& {Pfeiffer}}]{Gold2014b}
{Gold}, R., {Paschalidis}, V., {Ruiz}, M., {et~al.} 2014{\natexlab{b}}, \prd,
  90, 104030, \dodoi{10.1103/PhysRevD.90.104030}

\bibitem[{{Gole} {et~al.}(2016){Gole}, {Simon}, {Lubow}, \&
  {Armitage}}]{Gole2016}
{Gole}, D., {Simon}, J.~B., {Lubow}, S.~H., \& {Armitage}, P.~J. 2016, \apj,
  826, 18, \dodoi{10.3847/0004-637X/826/1/18}

\bibitem[{{Graham} {et~al.}(2015){Graham}, {Djorgovski}, {Stern}, {Glikman},
  {Drake}, {Mahabal}, {Donalek}, {Larson}, \&
  {Christensen}}]{2015Natur.518...74G}
{Graham}, M.~J., {Djorgovski}, S.~G., {Stern}, D., {et~al.} 2015, \nat, 518,
  74, \dodoi{10.1038/nature14143}

\bibitem[{{Hawley} {et~al.}(2011){Hawley}, {Guan}, \& {Krolik}}]{hgk11}
{Hawley}, J.~F., {Guan}, X., \& {Krolik}, J.~H. 2011, \apj, 738, 84,
  \dodoi{10.1088/0004-637X/738/1/84}

\bibitem[{{Hawley} \& {Krolik}(2002)}]{HK02}
{Hawley}, J.~F., \& {Krolik}, J.~H. 2002, \apj, 566, 164,
  \dodoi{10.1086/338059}

\bibitem[{{Heath} \& {Nixon}(2020)}]{HeathNixon2020}
{Heath}, R.~M., \& {Nixon}, C.~J. 2020, \aap, 641, A64,
  \dodoi{10.1051/0004-6361/202038548}

\bibitem[{Johnson-McDaniel {et~al.}(2009)Johnson-McDaniel, Yunes, Tichy, \&
  Owen}]{JohnsonMcDaniel:2009dq}
Johnson-McDaniel, N.~K., Yunes, N., Tichy, W., \& Owen, B.~J. 2009, Phys. Rev.,
  D80, 124039, \dodoi{10.1103/PhysRevD.80.124039}

\bibitem[{{Katz} {et~al.}(2020){Katz}, {Kelley}, {Dosopoulou}, {Berry},
  {Blecha}, \& {Larson}}]{Katz2020}
{Katz}, M.~L., {Kelley}, L.~Z., {Dosopoulou}, F., {et~al.} 2020, \mnras, 491,
  2301, \dodoi{10.1093/mnras/stz3102}

\bibitem[{{Kelley} {et~al.}(2019){Kelley}, {Charisi}, {Burke-Spolaor}, {Simon},
  {Blecha}, {Bogdanovic}, {Colpi}, {Comerford}, {D'Orazio}, {Dotti},
  {Eracleous}, {Graham}, {Greene}, {Haiman}, {Holley-Bockelmann}, {Kara},
  {Kelly}, {Komossa}, {Larson}, {Liu}, {Ma}, {Noble}, {Paschalidis}, {Rafikov},
  {Ravi}, {Runnoe}, {Sesana}, {Stern}, {Strauss}, {U}, {Volonteri}, \&
  {Nanograv Collaboration}}]{Kelley2019}
{Kelley}, L., {Charisi}, M., {Burke-Spolaor}, S., {et~al.} 2019, \baas, 51,
  490.
\newblock \doarXiv{1903.07644}

\bibitem[{{Keppler} {et~al.}(2020){Keppler}, {Penzlin}, {Benisty}, {van
  Boekel}, {Henning}, {van Holstein}, {Kley}, {Garufi}, {Ginski}, {Brandner},
  {Bertrang}, {Boccaletti}, {de Boer}, {Bonavita}, {Brown Sevilla}, {Chauvin},
  {Dominik}, {Janson}, {Langlois}, {Lodato}, {Maire}, {M{\'e}nard}, {Pantin},
  {Pinte}, {Stolker}, {Szul{\'a}gyi}, {Thebault}, {Villenave}, {Zurlo},
  {Rabou}, {Feautrier}, {Feldt}, {Madec}, \& {Wildi}}]{Keppler2020}
{Keppler}, M., {Penzlin}, A., {Benisty}, M., {et~al.} 2020, \aap, 639, A62,
  \dodoi{10.1051/0004-6361/202038032}

\bibitem[{{Klein} {et~al.}(2016){Klein}, {Barausse}, {Sesana}, {Petiteau},
  {Berti}, {Babak}, {Gair}, {Aoudia}, {Hinder}, {Ohme}, \&
  {Wardell}}]{Klein2016}
{Klein}, A., {Barausse}, E., {Sesana}, A., {et~al.} 2016, \prd, 93, 024003,
  \dodoi{10.1103/PhysRevD.93.024003}

\bibitem[{{Liao} {et~al.}(2021){Liao}, {Chen}, {Liu}, {Holgado}, {Guo},
  {Gruendl}, {Morganson}, {Shen}, {Davis}, {Kessler}, {Martini}, {McMahon},
  {Allam}, {Annis}, {Avila}, {Banerji}, {Bechtol}, {Bertin}, {Brooks},
  {Buckley-Geer}, {Carnero Rosell}, {Carrasco Kind}, {Carretero}, {Javier
  Castander}, {Cunha}, {D'Andrea}, {da Costa}, {Davis}, {De Vicente}, {Desai},
  {Thomas Diehl}, {Doel}, {Eifler}, {Evrard}, {Flaugher}, {Fosalba}, {Frieman},
  {Garcia-Bellido}, {Gaztanaga}, {Glazebrook}, {Gruen}, {Gschwend},
  {Gutierrez}, {Hartley}, {Hollowood}, {Honscheid}, {Hoyle}, {James}, {Krause},
  {Kuehn}, {Lima}, {Maia}, {Marshall}, {Menanteau}, {Miquel}, {Plazas
  Malag{\'o}n}, {Roodman}, {Sanchez}, {Scarpine}, {Schubnell}, {Serrano},
  {Smith}, {Smith}, {Soares-Santos}, {Sobreira}, {Suchyta}, {Swanson}, {Tarle},
  {Vikram}, \& {Walker}}]{Liao2021}
{Liao}, W.-T., {Chen}, Y.-C., {Liu}, X., {et~al.} 2021, \mnras, 500, 4025,
  \dodoi{10.1093/mnras/staa3055}

\bibitem[{{Liu} {et~al.}(2018){Liu}, {Gezari}, \& {Miller}}]{Liu2018}
{Liu}, T., {Gezari}, S., \& {Miller}, M.~C. 2018, \apjl, 859, L12,
  \dodoi{10.3847/2041-8213/aac2ed}

\bibitem[{{Liu} {et~al.}(2015){Liu}, {Gezari}, {Heinis}, {Magnier}, {Burgett},
  {Chambers}, {Flewelling}, {Huber}, {Hodapp}, {Kaiser}, {Kudritzki}, {Tonry},
  {Wainscoat}, \& {Waters}}]{2015ApJ...803L..16L}
{Liu}, T., {Gezari}, S., {Heinis}, S., {et~al.} 2015, \apjl, 803, L16,
  \dodoi{10.1088/2041-8205/803/2/L16}

\bibitem[{{Liu} {et~al.}(2019){Liu}, {Gezari}, {Ayers}, {Burgett}, {Chambers},
  {Hodapp}, {Huber}, {Kudritzki}, {Metcalfe}, {Tonry}, {Wainscoat}, \&
  {Waters}}]{Liu2019}
{Liu}, T., {Gezari}, S., {Ayers}, M., {et~al.} 2019, \apj, 884, 36,
  \dodoi{10.3847/1538-4357/ab40cb}

\bibitem[{{Lopez Armengol} {et~al.}(2021){Lopez Armengol}, {Combi},
  {Campanelli}, {Noble}, {Krolik}, {Bowen}, {Avara}, {Mewes}, \&
  {Nakano}}]{LopezArmengol2021}
{Lopez Armengol}, F.~G., {Combi}, L., {Campanelli}, M., {et~al.} 2021, arXiv
  e-prints, arXiv:2102.00243.
\newblock \doarXiv{2102.00243}

\bibitem[{{MacFadyen} \& {Milosavljevi{\'c}}(2008)}]{MM08}
{MacFadyen}, A.~I., \& {Milosavljevi{\'c}}, M. 2008, \apj, 672, 83,
  \dodoi{10.1086/523869}

\bibitem[{{Mangiagli} {et~al.}(2020){Mangiagli}, {Klein}, {Bonetti}, {Katz},
  {Sesana}, {Volonteri}, {Colpi}, {Marsat}, \& {Babak}}]{Mangiagli2020}
{Mangiagli}, A., {Klein}, A., {Bonetti}, M., {et~al.} 2020, \prd, 102, 084056,
  \dodoi{10.1103/PhysRevD.102.084056}

\bibitem[{{Miranda} {et~al.}(2017){Miranda}, {Mu{\~n}oz}, \&
  {Lai}}]{Miranda2017}
{Miranda}, R., {Mu{\~n}oz}, D.~J., \& {Lai}, D. 2017, \mnras, 466, 1170,
  \dodoi{10.1093/mnras/stw3189}

\bibitem[{{Moody} {et~al.}(2019){Moody}, {Shi}, \& {Stone}}]{Moody2019}
{Moody}, M. S.~L., {Shi}, J.-M., \& {Stone}, J.~M. 2019, \apj, 875, 66,
  \dodoi{10.3847/1538-4357/ab09ee}

\bibitem[{{M{\"o}sta} {et~al.}(2019){M{\"o}sta}, {Taam}, \&
  {Duffell}}]{Mosta2019}
{M{\"o}sta}, P., {Taam}, R.~E., \& {Duffell}, P.~C. 2019, \apjl, 875, L21,
  \dodoi{10.3847/2041-8213/ab1592}

\bibitem[{{Mu{\~n}oz} \& {Lai}(2016)}]{Munoz2016}
{Mu{\~n}oz}, D.~J., \& {Lai}, D. 2016, \apj, 827, 43,
  \dodoi{10.3847/0004-637X/827/1/43}

\bibitem[{{Mu{\~n}oz} {et~al.}(2020){Mu{\~n}oz}, {Lai}, {Kratter}, \&
  {Miranda}}]{Munoz2020a}
{Mu{\~n}oz}, D.~J., {Lai}, D., {Kratter}, K., \& {Miranda}, R. 2020, \apj, 889,
  114, \dodoi{10.3847/1538-4357/ab5d33}

\bibitem[{{Mu{\~n}oz} \& {Lithwick}(2020)}]{Munoz2020b}
{Mu{\~n}oz}, D.~J., \& {Lithwick}, Y. 2020, \apj, 905, 106,
  \dodoi{10.3847/1538-4357/abc74c}

\bibitem[{{Mu{\~n}oz} {et~al.}(2019){Mu{\~n}oz}, {Miranda}, \&
  {Lai}}]{Munoz2019}
{Mu{\~n}oz}, D.~J., {Miranda}, R., \& {Lai}, D. 2019, \apj, 871, 84,
  \dodoi{10.3847/1538-4357/aaf867}

\bibitem[{Mundim {et~al.}(2014)Mundim, Nakano, Yunes, Campanelli, Noble, \&
  Zlochower}]{Mundim:2013vca}
Mundim, B.~C., Nakano, H., Yunes, N., {et~al.} 2014, Phys. Rev., D89, 084008,
  \dodoi{10.1103/PhysRevD.89.084008}

\bibitem[{{Noble} {et~al.}(2006){Noble}, {Gammie}, {McKinney}, \& {Del
  Zanna}}]{Noble06}
{Noble}, S.~C., {Gammie}, C.~F., {McKinney}, J.~C., \& {Del Zanna}, L. 2006,
  \apj, 641, 626, \dodoi{10.1086/500349}

\bibitem[{{Noble} \& {Krolik}(2009)}]{NK09}
{Noble}, S.~C., \& {Krolik}, J.~H. 2009, \apj, 703, 964,
  \dodoi{10.1088/0004-637X/703/1/964}

\bibitem[{{Noble} {et~al.}(2009){Noble}, {Krolik}, \& {Hawley}}]{Noble09}
{Noble}, S.~C., {Krolik}, J.~H., \& {Hawley}, J.~F. 2009, \apj, 692, 411,
  \dodoi{10.1088/0004-637X/692/1/411}

\bibitem[{{Noble} {et~al.}(2010){Noble}, {Krolik}, \& {Hawley}}]{Noble10}
---. 2010, \apj, 711, 959, \dodoi{10.1088/0004-637X/711/2/959}

\bibitem[{{Noble} {et~al.}(2012){Noble}, {Mundim}, {Nakano}, {Krolik},
  {Campanelli}, {Zlochower}, \& {Yunes}}]{Noble12}
{Noble}, S.~C., {Mundim}, B.~C., {Nakano}, H., {et~al.} 2012, \apj, 755, 51,
  \dodoi{10.1088/0004-637X/755/1/51}

\bibitem[{{Paschalidis} {et~al.}(2021){Paschalidis}, {Bright}, {Ruiz}, \&
  {Gold}}]{Paschalidis2021}
{Paschalidis}, V., {Bright}, J., {Ruiz}, M., \& {Gold}, R. 2021, arXiv
  e-prints, arXiv:2102.06712.
\newblock \doarXiv{2102.06712}

\bibitem[{{Ragusa} {et~al.}(2020){Ragusa}, {Alexander}, {Calcino}, {Hirsh}, \&
  {Price}}]{Ragusa2020}
{Ragusa}, E., {Alexander}, R., {Calcino}, J., {Hirsh}, K., \& {Price}, D.~J.
  2020, \mnras, 499, 3362, \dodoi{10.1093/mnras/staa2954}

\bibitem[{{Ragusa} {et~al.}(2017){Ragusa}, {Dipierro}, {Lodato}, {Laibe}, \&
  {Price}}]{Ragusa2017}
{Ragusa}, E., {Dipierro}, G., {Lodato}, G., {Laibe}, G., \& {Price}, D.~J.
  2017, \mnras, 464, 1449, \dodoi{10.1093/mnras/stw2456}

\bibitem[{{Ragusa} {et~al.}(2016){Ragusa}, {Lodato}, \& {Price}}]{Ragusa2016}
{Ragusa}, E., {Lodato}, G., \& {Price}, D.~J. 2016, \mnras, 460, 1243,
  \dodoi{10.1093/mnras/stw1081}

\bibitem[{{Roedig} {et~al.}(2014){Roedig}, {Krolik}, \& {Miller}}]{RoedigKM14}
{Roedig}, C., {Krolik}, J.~H., \& {Miller}, M.~C. 2014, \apj, 785, 115,
  \dodoi{10.1088/0004-637X/785/2/115}

\bibitem[{Sago {et~al.}(2016)Sago, Fujita, \& Nakano}]{Sago:2016xsp}
Sago, N., Fujita, R., \& Nakano, H. 2016, Phys. Rev. D, 93, 104023,
  \dodoi{10.1103/PhysRevD.93.104023}

\bibitem[{{Sano} {et~al.}(2004){Sano}, {Inutsuka}, {Turner}, \&
  {Stone}}]{2004ApJ...605..321S}
{Sano}, T., {Inutsuka}, S.-i., {Turner}, N.~J., \& {Stone}, J.~M. 2004, \apj,
  605, 321, \dodoi{10.1086/382184}

\bibitem[{{Sesana} {et~al.}(2012){Sesana}, {Roedig}, {Reynolds}, \&
  {Dotti}}]{Sesana2012}
{Sesana}, A., {Roedig}, C., {Reynolds}, M.~T., \& {Dotti}, M. 2012, \mnras,
  420, 860, \dodoi{10.1111/j.1365-2966.2011.20097.x}

\bibitem[{{Shi} \& {Krolik}(2015)}]{Shi15}
{Shi}, J.-M., \& {Krolik}, J.~H. 2015, \apj, 807, 131,
  \dodoi{10.1088/0004-637X/807/2/131}

\bibitem[{{Shi} \& {Krolik}(2016)}]{Shi2016}
---. 2016, \apj, 832, 22, \dodoi{10.3847/0004-637X/832/1/22}

\bibitem[{{Shi} {et~al.}(2012){Shi}, {Krolik}, {Lubow}, \& {Hawley}}]{Shi12}
{Shi}, J.-M., {Krolik}, J.~H., {Lubow}, S.~H., \& {Hawley}, J.~F. 2012, \apj,
  749, 118, \dodoi{10.1088/0004-637X/749/2/118}

\bibitem[{{Sorathia} {et~al.}(2012){Sorathia}, {Reynolds}, {Stone}, \&
  {Beckwith}}]{2012ApJ...749..189S}
{Sorathia}, K.~A., {Reynolds}, C.~S., {Stone}, J.~M., \& {Beckwith}, K. 2012,
  \apj, 749, 189, \dodoi{10.1088/0004-637X/749/2/189}

\bibitem[{Szilagyi {et~al.}(2015)Szilagyi, Blackman, Buonanno, Taracchini,
  Pfeiffer, Scheel, Chu, Kidder, \& Pan}]{Szilagyi:2015rwa}
Szilagyi, B., Blackman, J., Buonanno, A., {et~al.} 2015, Phys. Rev. Lett., 115,
  031102, \dodoi{10.1103/PhysRevLett.115.031102}

\bibitem[{{Tiede} {et~al.}(2020){Tiede}, {Zrake}, {MacFadyen}, \&
  {Haiman}}]{Tiede2020}
{Tiede}, C., {Zrake}, J., {MacFadyen}, A., \& {Haiman}, Z. 2020, \apj, 900, 43,
  \dodoi{10.3847/1538-4357/aba432}

\bibitem[{{T{\'o}th}(2000)}]{Toth00}
{T{\'o}th}, G. 2000, Journal of Computational Physics, 161, 605

\bibitem[{Yunes \& Berti(2008)}]{Yunes:2008tw}
Yunes, N., \& Berti, E. 2008, Phys. Rev., D77, 124006,
  \dodoi{10.1103/PhysRevD.77.124006, 10.1103/PhysRevD.83.109901,
  10.1103/PhysRevD.77.124006, 10.1103/PhysRevD.83.109901}

\bibitem[{Yunes \& Tichy(2006)}]{Yunes:2006iw}
Yunes, N., \& Tichy, W. 2006, Phys. Rev., D74, 064013,
  \dodoi{10.1103/PhysRevD.74.064013}

\bibitem[{Yunes {et~al.}(2006)Yunes, Tichy, Owen, \& Bruegmann}]{Yunes:2005nn}
Yunes, N., Tichy, W., Owen, B.~J., \& Bruegmann, B. 2006, Phys. Rev., D74,
  104011, \dodoi{10.1103/PhysRevD.74.104011}

\bibitem[{Zhang {et~al.}(2011)Zhang, Yunes, \& Berti}]{Zhang:2011vha}
Zhang, Z., Yunes, N., \& Berti, E. 2011, Phys. Rev., D84, 024029,
  \dodoi{10.1103/PhysRevD.84.024029}

\bibitem[{{Zhu} \& {Thrane}(2020)}]{ZhuThrane2020}
{Zhu}, X.-J., \& {Thrane}, E. 2020, \apj, 900, 117,
  \dodoi{10.3847/1538-4357/abac5a}

\bibitem[{{Zilh{\~a}o} {et~al.}(2015){Zilh{\~a}o}, {Noble}, {Campanelli}, \&
  {Zlochower}}]{Zilhao2015}
{Zilh{\~a}o}, M., {Noble}, S.~C., {Campanelli}, M., \& {Zlochower}, Y. 2015,
  \prd, 91, 024034, \dodoi{10.1103/PhysRevD.91.024034}

\bibitem[{{Zrake} {et~al.}(2020){Zrake}, {Tiede}, {MacFadyen}, \&
  {Haiman}}]{Zrake2020}
{Zrake}, J., {Tiede}, C., {MacFadyen}, A., \& {Haiman}, Z. 2020, arXiv
  e-prints, arXiv:2010.09707.
\newblock \doarXiv{2010.09707}

\end{thebibliography}
\bibliographystyle{aasjournal}

\end{document}